\documentclass[utf8]{FrontiersinHarvard} % for articles in journals using the Harvard Referencing Style (Author-Date), for Frontiers Reference Styles by Journal: https://zendesk.frontiersin.org/hc/en-us/articles/360017860337-Frontiers-Reference-Styles-by-Journal
\usepackage{url,hyperref,lineno,microtype,subcaption}
\usepackage[onehalfspacing]{setspace}
\usepackage{amsmath}
\usepackage[T1]{fontenc}
\usepackage{aas_macros}
\def \arcsec {\hbox{$^{\prime\prime}$}}
\def\keyFont{\fontsize{8}{11}\helveticabold }
\def\firstAuthorLast{Shen {et~al.}} %use et al only if is more than 1 author
\def\Authors{Chengcai Shen\,$^{1,*}$, Vanessa Polito\,$^{2,3,4}$, Katharine K. Reeves\,$^{1}$, Bin Chen\,$^{5}$,  Sijie Yu\,$^{5}$, and Xiaoyan Xie\,$^{1}$}
\def\Address{$^{1}$Center for Astrophysics $\mid$ Harvard \& Smithsonian, Cambridge, MA, USA \\
$^{2}$Bay Area Environmental Research Institute, NASA Research Park, Moffett Field, CA, USA \\
$^{3}$Lockheed Martin Solar and Astrophysics Laboratory, Palo Alto, CA, USA \\
$^{4}$Department of Physics, Oregon State University, 301 Weniger Hall, Corvallis, OR 97331\\
$^{5}$Center for Solar-Terrestrial Research, New Jersey Institute of Technology, University Heights, Newark, NJ, USA}
\def\corrAuthor{Corresponding Author}

\def\corrEmail{chengcaishen@cfa.harvard.edu}

\begin{document}
\onecolumn
\firstpage{1}

%\title[Plasma Emission Variations in 3D models]{Plasma Emission Variations Due to Dynamic Structures in Reconnection Regions During Solar Eruptions} 
%\title[Plasma Emission Variations in 3D models]{A Model of Plasma Emission Variations in Magnetic Reconnection Regions During Solar Eruptions} 
\title[Non-thermal Broadening in MR]{Non-thermal Broadening of IRIS Fe XXI Lines Caused by Turbulent Plasma Flows in the Magnetic Reconnection Region During Solar Eruptions}

\author[\firstAuthorLast ]{\Authors} %This field will be automatically populated
\address{} %This field will be automatically populated
\correspondance{} %This field will be automatically populated

\extraAuth{}% If there are more than 1 corresponding author, comment this line and uncomment the next one.
%\extraAuth{corresponding Author2 \\ Laboratory X2, Institute X2, Department X2, Organization X2, Street X2, City X2 , State XX2 (only USA, Canada and Australia), Zip Code2, X2 Country X2, email2@uni2.edu}

\maketitle

\begin{abstract}

%%% Leave the Abstract empty if your article does not require one, please see the Summary Table for full details.

Magnetic reconnection is the key mechanism for energy release in solar eruptions, where the high-temperature emission is the primary diagnostic for investigating the plasma properties during the reconnection process. Non-thermal broadening of high-temperature lines has been observed in both the reconnection current sheet (CS) and flare loop-top regions by UV spectrometers, but its origin remains unclear. 
In this work, we use a recently developed three-dimensional magnetohydrodynamic (MHD) simulation to model magnetic reconnection in the standard solar flare geometry and reveal highly dynamic plasma flows in the reconnection regions. 
We calculate the synthetic profiles of the Fe XXI 1354 \AA~line observed by the Interface Region Imaging Spectrograph (IRIS) spacecraft by using parameters of the MHD model, including plasma density, temperature, and velocity. 
Our model shows that the turbulent bulk plasma flows in the CS and flare loop-top regions are responsible for the non-thermal broadening of the Fe XXI emission line. The modeled non-thermal velocity ranges from tens of km s$^{-1}$ to more than two hundred km s$^{-1}$, which is consistent with the IRIS observations. 
Simulated  2D spectral line maps around the reconnection region also  reveal  highly dynamic downwflow structures where the  high non-thermal velocity is large, which is consistent with the observations as well.

\tiny
 \keyFont{ \section{Keywords:} Solar corona, Spectroscopic, Magnetohydrodynamical simulations} %All article types: you may provide up to 8 keywords; at least 5 are mandatory.
\end{abstract}

\section{Introduction}
%% P1: Magnetic and turbulent flows: what is? how does it appear? 
Magnetic reconnection, or the breaking and rejoining of magnetic field lines in a highly conducting plasma, is commonly believed to be a fundamental process during solar eruptions, and it plays a key role in rapid magnetic energy release \citep[e.g.,][]{Shibata2011LRSP....8....6S}. 
During a solar eruption, the magnetic energy can be quickly transported to kinetic and thermal energy in the reconnection diffusion and exhausting regions, and generate energetic particles.
The spectroscopic investigations of ultraviolet (UV) emission lines serve as an important diagnostic tool because these diagnostics include physical information about the high-temperature plasma, such as the dynamic evolution properties and the heating and particle acceleration mechanisms. Various parameters, such as temperature, thickness, density, turbulent velocity, and reconnection rate are interpreted from the observations of the CS. 
One of the most interesting features is the broadening of high temperature emission lines formed during flares .%, which is formed directly by multiple emission shifts on the wavelength. 
The theoretical explanation for such broadening includes microscopic ion motions and macroscopic plasma motions of the emitting ion. 
Therefore, it is important to investigate the impacts and contribution of the turbulent flows on observable emission lines to understand the reconnection process during solar eruptions.
Recently, high-temperature plasma due to magnetic reconnection has been widely observed in UV and Extreme-UV (EUV) emission lines, including observations of the entire reconnection site from the large scale-reconnection current sheet down to the flare loop-top region.

% P2: spectroscopic obs in current sheets and flare loop-tops regions?
In post coronal mass ejection (CME) plasma sheets, the spectroscopic observations from the Ultraviolet Coronagraph Spectrometer (UVCS) on the Solar and Heliospheric Observatory (SoHO) are well studied from large heights above the solar surface \citep[e.g., reviewed in][]{Lin2015SSRv..194..237L}. Among them, turbulent motions inside these hot plasma sheets have been reported by many authors based on the observations of UV/EUV spectral lines. For instance, \cite{Ciaravella2002ApJ...575.1116C} deduced that the turbulent motion had a speed less than 60 km s$^{-1}$ in the CS; \cite{Ciaravella2008ApJ...686.1372C} reported large non-thermal [Fe xviii] line width in the CS, which at the early stage of the CS is as high as 380 km s$^{-1}$ and later ranges between 50 and 200 km s$^{-1}$. The combination of bulk motions (e.g., the flow along a fan-shaped CS if the fan is seen edge-on) and turbulence \citep[e.g.,][]{Lazarian1999ApJ...517..700L} are suggested to be the main contributions to the line broadening.
In a long duration, CS studied by \cite{Bemporad2008ApJ...689..572B}, the derived non-thermal speeds in the Fe xviii 974.8 \AA spectral lines are of the order of $\sim$60 km s$^{-1}$ a few hours after the CME and slowly decay down to about 30 km s$^{-1}$ in the following 2 days.

At lower heights of CME/flare plasma sheets (up to about 1.15 R$_{sun}$ from the center of Sun), the EUV Imaging Spectrometer (EIS) on Hinode \citep{Kosugi2007SoPh..243....3K} has reported a large amount of spectroscopic observations of the plasma sheet region. 
\cite{Warren2018ApJ...854..122W} analyzed the EIS Fe XXIV lines and found that strong broadening of the 192.04 \AA~line occurs at the largest observed heights along the CS in a large X-class CME/flare eruption on 2017 September 10th. For instance, the measured non-thermal velocity rises from about 87 km s$^{-1}$ to about 152 km s$^{-1}$ over the observed length of the current sheet in a short time period (the 16:09 UT raster). Furthermore, the line broadening is strong very early in the flare and diminishes over time. 
Large non-thermal velocities up to 200 km s$^{-1}$ in EIS Fe XXIV line are also analyzed by \cite{Li2018ApJ...853L..15L} during this eruption.
In addition, the presence of non-Maxwellian electron distributions with enhanced high-energy tails has also been reported by \cite{Polito2018ApJ...864...63P} during the impulsive phase of this event. The observed line widths of the Fe XXIII and Fe XXIV EIS lines still imply considerable non-thermal broadening in excess of $\sim$200 km s$^{-1}$ above the flare loop-top region.

The non-thermal features are also be found in other high-temperature Fe XXI lines (e.g., 11MK). Using the Interface Region Imaging Spectrograph \citep[IRIS][]{DePontieu2014},  \cite{Tian2014ApJ...797L..14T} reported the red-shifted features of Fe XXI that coincided with an X-ray source as observed by RHESSI above flare loops. They obtained greatly redshifted velocities ($\sim$125 km s$^{-1}$ along the line of sight, LOS) on the Fe XXI 1354.08 \AA\ emission line with large non-thermal widths ($\sim$100 km s$^{-1}$) at the reconnection region.

%{\color{magenta}{it feels like we are jumping from IRIS to EIS and back to IRIS?}}
In recent flare fan observations (seen face\-on),  the non-thermal widths of the Fe XXI line observed by IRIS were also measured by \cite{Reeves2020ApJ...905..165R}.
In this work, they found that the pixels at the fan top have broader non-thermal widths than the rest of the emission, especially for the later rasters. By comparing the spectral results of the IRIS and Hinode/EIS observations with the synthetic results of 2D MHD simulations, \cite{Cai2022ApJ...929...99C} suggested that a compressed interface with apparent changes in intensity and Doppler velocities of the spectral lines exist above the flare loops, possibly related to termination shocks.
The temporal variation of non-thermal velocity was also studied in flare fan regions. \cite{Cai2019MNRAS.489.3183C} showed that the non-thermal velocity ranged from 19 to 64 km s$^{-1}$ during a particular period (between 16:00 UT and 16:26 UT, on 2017 September 10th) by analyzing the IRIS Fe XXI 1354.08 \AA\ spectral lines. It is worth mentioning that the observed line width in their work decreased with time as well, though the time period ($\sim$ half an hour) was shorter than these studies in long durations post-CME current sheets \citep[e.g.,][]{Bemporad2008ApJ...689..572B}.
The spatial variation of non-thermal velocities is reported by \cite{Doschek2014ApJ...788...26D} in which, they found that the non-thermal motions in the multimillion-degree regions increase with height above flare loops. The deduced non-thermal velocity ranges from 40 to 60 km s$^{-1}$ in the EIS Fe XXIII and Fe XXIV lines spectra.

%P3: Our recent model
The broadening of spectral line profiles is commonly thought to be the result mainly due to turbulent and bulk flows in solar flares.
In theory, the plasma sheet and flare loop-top regions (which are generally referenced as flare fans when viewed face-on) are predicted to contain substantial turbulence in a set of theoretical models. For example, the tearing current sheet and multiple plasma instabilities (e.g., the plasmoid instability) may cause a set of small-scale turbulent structures which has been widely found in recent numerical models \citep[e.g.,][]{Huang2016ApJ...818...20H,Ye2020ApJ...897...64Y}. 
Once the Alfv\'{e}nic reconnection downflows have collided with the closed magnetic loops, the dramatic variation of plasma $\beta$ also creates favorable environments for generating turbulent flows in these interface regions, where the $\beta$ can change from high-beta values in the current sheet region to low-beta states in the potential flare loops \citep[e.g.,][]{Shen2022NatAs...6..317S}.
Bulk plasma flows may significantly contribute to line profiles as well \citep{Ciaravella2008ApJ...686.1372C}. 
For instance, \cite{Guo2017ApJ...846L..12G} calculated synthetic spectral line profiles of the IRIS Fe XXI 1354 \AA\ line at the reconnection site with the presence of termination shocks (TS) in an MHD model of a solar flare. The significant shifts of these synthetic IRIS spectral lines suggested that the synthetic line profile of Fe XXI and its time evolution may serve as a possible guide for observational signatures of a flare TS. Possible Doppler shift signatures of such termination shocks have been subsequently reported by \citet{Polito2018b} using IRIS Fe XXI observations.
Further, in a comprehensive radiative 3D model of solar flare \citep{Cheung2019NatAs...3..160C}, synthetic non-Gaussian ($\kappa$) distributions of EUV emission-line profiles result from temperature and velocity gradients along the line of sight.

%P4: Paper structures: how does the turbulent flow contribute the EUV emission line profiles
In observations, various dynamic flows at multiple scales have also been found in a set of image observations above flare loops. For instance, \cite{McKenzie2013ApJ...766...39M} reported turbulent dynamic flows observed by the X-ray Telescope \citep[XRT;][]{Golub2007} on board Hinode and the Atmospheric Image Assembly \citep[AIA;][]{Lemen12} on board the solar dynamic observatory \citep[SDO;][]{Pesnell12}. The flows were found above the post-eruption arcades and measured with local correlation tracking. These results also show significant shears in velocity, giving the appearance of vortices and stagnations. More plasma downflows, now referred to as supra-arcade downflows (SADs), have often been investigated in the flare fan regions above the post-flare loops \citep[e.g.,][]{Xie2022ApJ...933...15X}.
However, studies focusing on how  these turbulent flows affect the EUV line profiles and the non-thermal broad features above the post-flare loops remain sparse. 

In this study, we focus on the properties of Fe XXI lines in the plasma sheet and flare-loop top regions.
The high-temperature plasma in flare cusp regions is well observed in the 2015-03-07 flare by IRIS, Hinode/XRT, and SDO/AIA.
In Section~\ref{Sect:methods}, we will briefly review the main observational features and show the non-thermal broadening and Doppler velocity variations observed by the Fe XXI line.
On the modeling side, in our recent state-of-the-art MHD model of solar flares \cite{Shen2022NatAs...6..317S}, a turbulent interface region below the flare termination shock is revealed in the classic solar flare geometry. The highly dynamic flows are also found in the CS regions. 
Therefore, we aim to investigate the emission lines based on this 3D model and discuss the origin of line broadening to be compared with observations.
We describe the numerical models and the calculation method for simulating Fe XXI lines in Section~\ref{Sect:methods}, and show the synthetic spectral lines in Section ~\ref{Sect:results}. 
We then investigate the spatial and temporal distribution of the non-thermal spectral line broadening by combining 3D MHD modeling and IRIS spectral observations. Finally, discussions and conclusions are given in Section~\ref{Sect:discussion}.

\begin{figure}[h!]
    \begin{center}
    \includegraphics[width=0.7\textwidth]{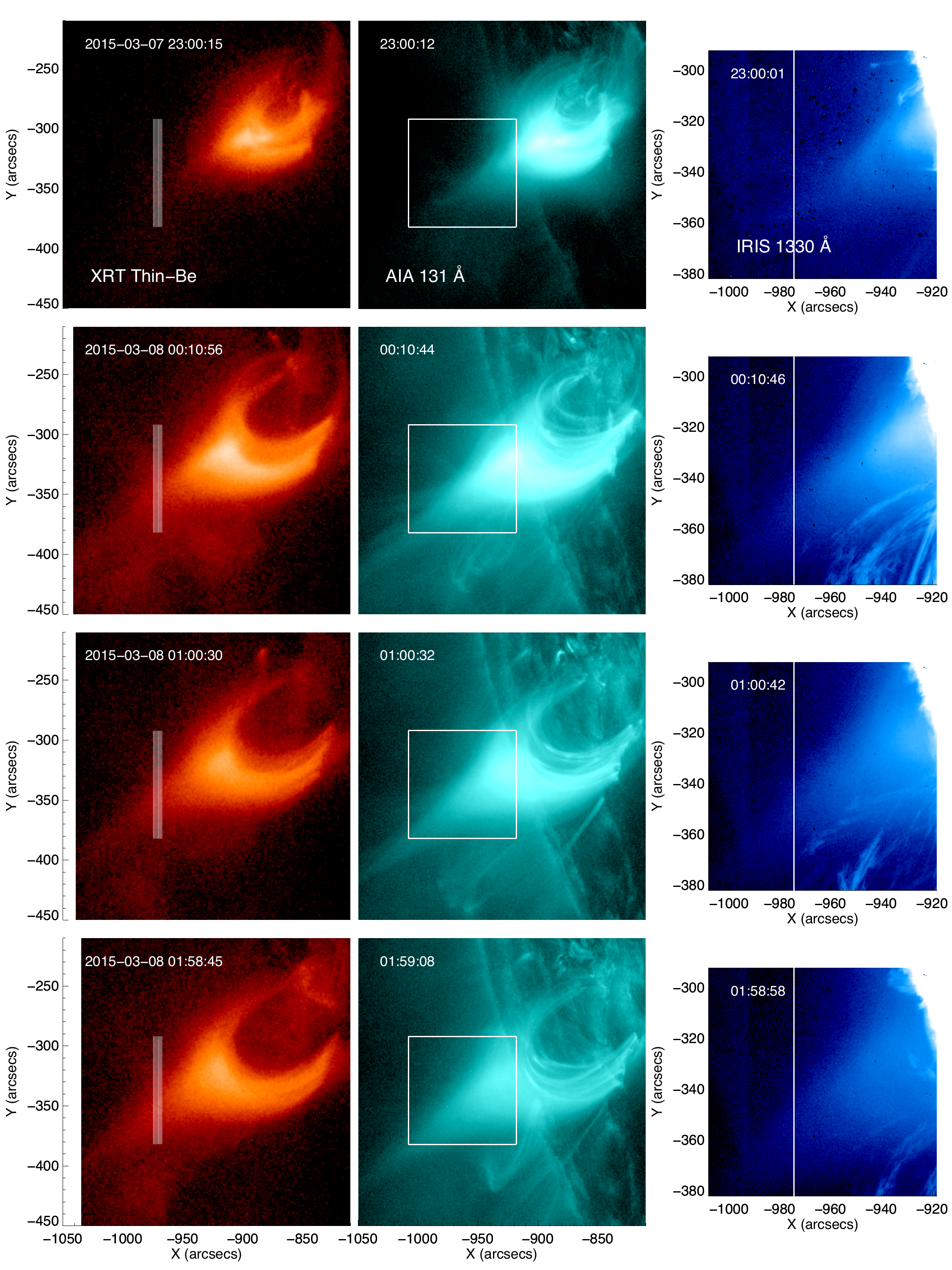}
    \end{center}
    \caption{Overview of the 2015-03-07 flare between 23:00:15 UT to 01:58:45 UT on March 8th. The panels from left to right are observed images from \textit{Hinode}/XRT, \textit{SDO}/AIA 131\AA, and \textit{IRIS} 1330 \AA\ SJI. The vertical white lines on XRT images indicate the slit position of \textit{IRIS} Fe XXI spectra observations. The white box on AIA 131 maps shows the SJI FOV.}
    \label{fig:obs_overview}
\end{figure}

\section{Material and Methods}
\label{Sect:methods}
% For Original Research articles, please note that the Material and Methods section can be placed in any of the following ways: before Results, before Discussion or after Discussion.

\begin{figure}[h!]
    \begin{center}
    \includegraphics[width=1\textwidth]{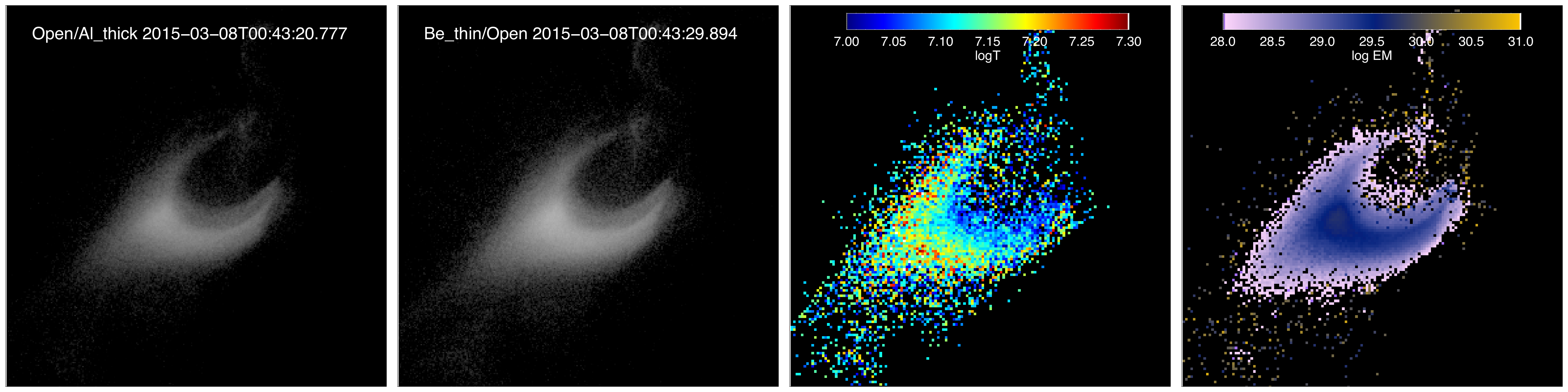}
    \end{center}
    \caption{Left two images: XRT Al-Thick and Be-Thin, respectively, at about 00:43 UT on March 8.  Right two images: Temperature and emission measure, respectively, calculated from the Al-Thick/Be-Thin filter ratio.}
    \label{fig:xrt_teem}
\end{figure}

\begin{figure*}[t!]
    \centering
    \includegraphics[width=0.7\textwidth]{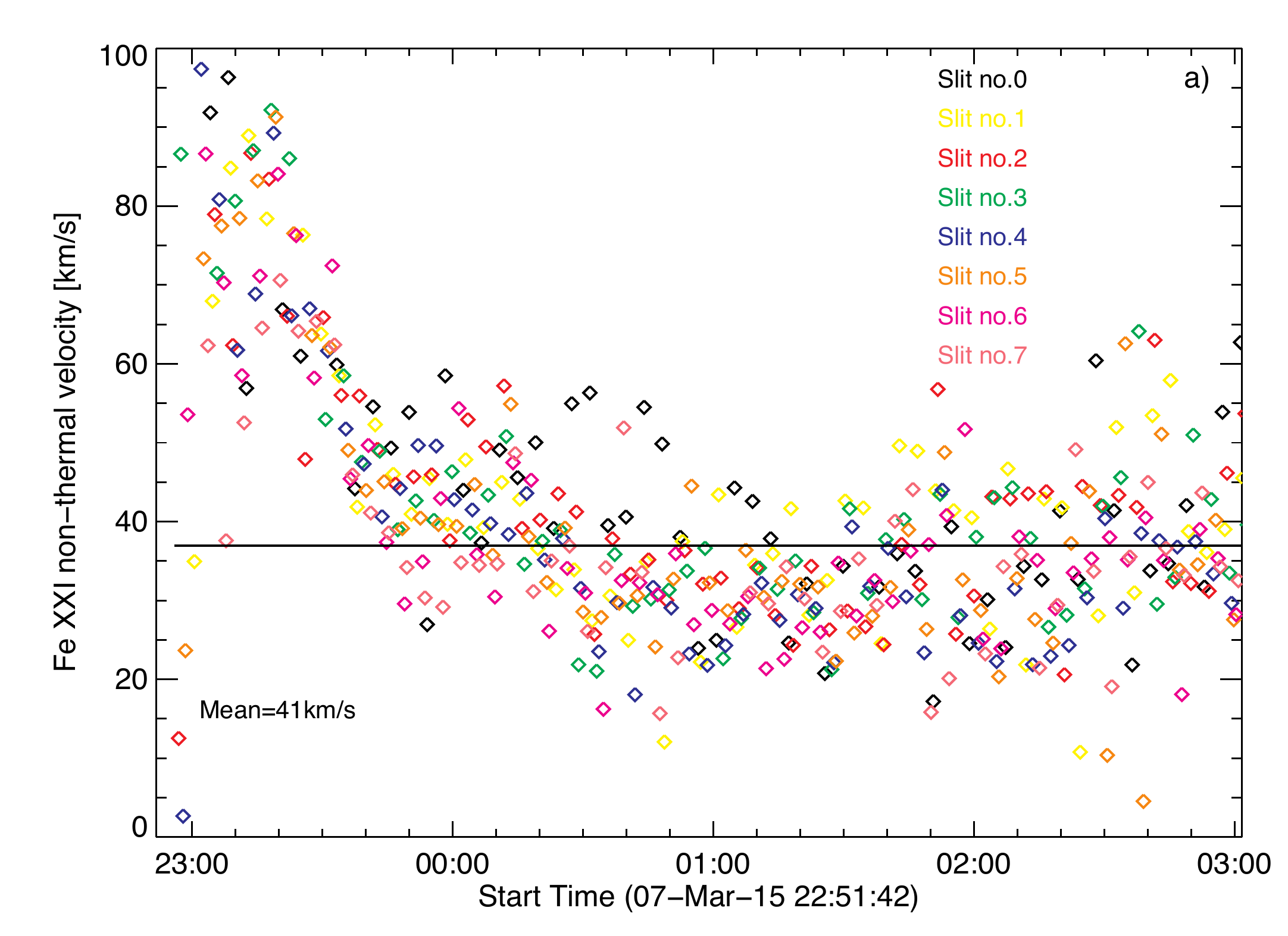}
    \includegraphics[width=0.7\textwidth]{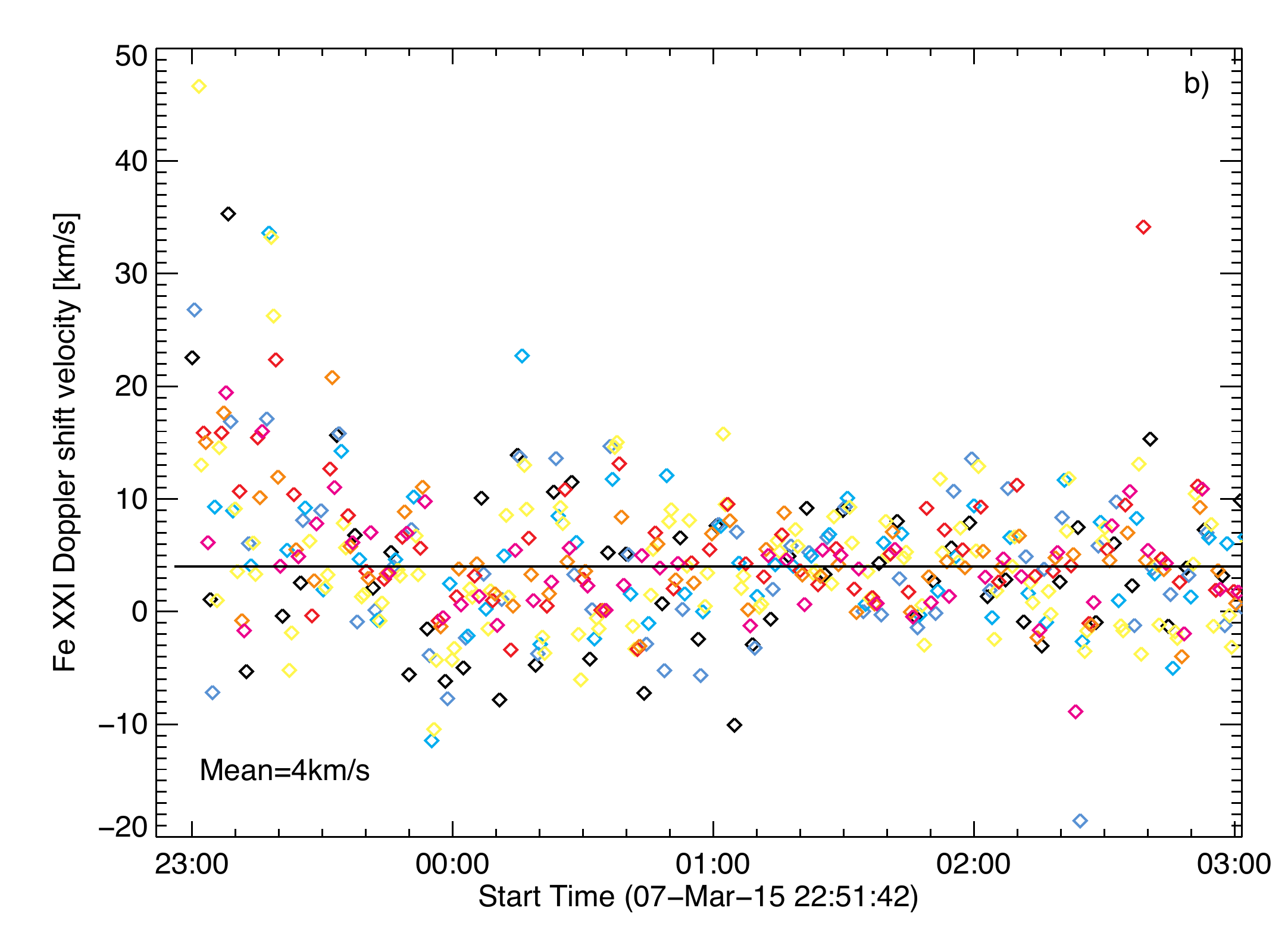}
    \caption{Panel (a): Non-thermal velocity  of the Fe XXI line observed by the 8 IRIS raster slit overtime during the 2015-03-07 flare. Panel (b): Fe XXI velocity over time for the same event.}
    \label{fig:fexxi}
\end{figure*}

\subsection{Observations}
The flare event under study is a long-duration M9.2 solar flare located at the east limb, that occurred on March 7, 2015.  We mostly focus on observations from the IRIS satellite, which, since its launch in 2013, has been providing unprecedented high-resolution images and spectra of the lower atmosphere and corona \citep{DePontieu2014,DePontieu2021}. IRIS consists of: (1) a Slit-Jaw Imager (SJI) channel, acquiring images in four different filters showing plasma at photospheric, chromospheric and transition region temperatures with a 0.167\arcsec~resolution; and (2) a spectrograph channel, observing emission lines and continua formed over a broad range of temperatures (from the photosphere to the flaring corona), at very high spatial (0.33--0.4\arcsec), temporal (down to $\approx$~1s) and spectral (2.7 km s$^{-1}$ pixels) resolution.  The level 2 IRIS data used here are already calibrated and prepped, as described in the IRIS' documentation and instrument papers \citep{DePontieu2014, Wuelser2018}. 

The {\it Hinode} X-Ray Telescope \citep{Golub2007,Kano2008} observed the flare with its flare response, using the Be-Thin and Be-Med filters and a resolution of 1.028\arcsec\ per pixel.  A two-filter observing program consisting of the Al-poly and Be-thin filters was taken before and after the flare response.  The XRT data is calibrated and prepped using the {\tt xrt\_prep} routine, available in the SolarSoft suite of IDL programs \citep{Freeland1998}, which does the dark subtraction, and removes the pedestal and vignetting effects \citep{Kobelski2014}.  The current version of {\tt xrt\_prep} also aligns the XRT images with AIA images by calibrating the time-dependent offsets between XRT and the {\it Hinode} Ultra Fine Sun Sensors \citep{Yoshimura2015}. 

AIA \citep{Lemen12} provides full-Sun images with a resolution of $\sim$0.6\arcsec\ per pixel and a cadence of 12 seconds for the EUV passbands.  The AIA data are processed using the SolarSoft routine {\tt aia\_prep}, which de-rotates the images from the different AIA telescopes, aligns them, and gives them all the same plate scale.  

%{\color{magenta}{Kathy can you check that the description of your figure below is accurate?}}. 
%KKR: I checked it and made some edits.
Figure \ref{fig:obs_overview} shows an overview of the flare as observed by XRT in the Be-Thin filter (left columns), AIA in the 131\AA~filter (dominated by Fe XXI emission at $\approx$~10MK, middle column) and in the IRIS SJI centered at 1330\AA\ (right column).  This SJI channel is dominated by emission from C II plasma at around 10--40kK, but it is also sensitive to hot plasma from Fe XXI during flares. The vertical lines in the XRT images indicate the location of the IRIS spectrograph slit during the observation. The white box in the middle images indicates the field of view of the IRIS SJI images on the right. 

Figure \ref{fig:xrt_teem} shows two images from the XRT flare response using the Al-Thick and Be-Thin filters.  The right two panels of Figure \ref{fig:xrt_teem} shows the temperature and emission measure, calculated from the filter ratio of the XRT two filters using the IDL routine {\tt xrt\_teem\_ch}.  We use coronal abundances to calculate the XRT temperature response functions because we are interested in the above-the-looptop plasma, which is likely directly heated, instead of resulting from chromospheric evaporation.  Spectroscopic measurements in this region during another flare indicate that abundances are likely coronal \citep{Warren2018ApJ...854..122W}.

In this work, we focus on the analysis of the Fe XXI observed by IRIS in high-temperature $\approx$~10MK plasma during flares. The IRIS observations during the flare event under study consist of an 8-step sparse raster (where the distance between consecutive slit positions is 1\arcsec), with a $\approx$~30s exposure time, a raster cadence of $\approx$~250s and a factor of 2 summing along the slit position. Figure~\ref{fig:fexxi} shows the non-thermal velocity (a) and Doppler shift velocity (b) of the Fe XXI line for the 8 slit positions as a function of time. For each slit position and time we averaged the Fe XXI spectra over 50 slit pixels to increase the signal-to-noise ratio. In fact, the line is very weak in the cusp/reconnection region, which is not surprising given the low density of the plasma there and the fact that the IRIS Fe XXI 1354.08\AA~line is a forbidden transition \citep[e.g.][]{Young2015}. The non-thermal velocity was obtained by using the following formula:

\begin{equation}
    v_{nth}=\frac{\lambda_0}{c}\cdot~2\sqrt{ln2}~\cdot~\sqrt{FWHM^2-FWHM_{th}^2-FWHM_{instr}^2}\cdot~
\end{equation}
where $\lambda_0$ is 1354.08\AA, the rest wavelength of the line \citep[e.g.][]{Polito2015}, $c$ is 3$\times$10$^{10}$cm s$^{-1}$, $FWHM_{instr}$ is the IRIS instrumental width (26m\AA) and $FWHM_{th}$ is the thermal broadening of the line ($\approx$~0.43\AA) assuming a formation temperature of log$T$[K]=7.05.

Figure~\ref{fig:fexxi}a) shows that the non-thermal velocity gradually decreases over time from more than 100 km s$^{-1}$ to about 30~km s$^{-1}$ in around two hours. On the other hand, Figure~\ref{fig:fexxi}b) shows that the Doppler shift velocity does not have a similar trend but appears to be almost constant around a mean value of 4 km s$^{-1}$ for most of the time. The only exception is the first few minutes of the observation, where the Doppler velocity reaches larger values up to almost 40 km s$^{-1}$. We note that there is a lot of scattering between the values obtained in different IRIS slits.  The IRIS level 2 data used in this work are already corrected for both the orbital and absolute calibration of the wavelength array,  with an accuracy that is estimated to be $\approx$~1 km s$^{-1}$ for most databases \citep{Wuelser2018}. However, it is usually recommended to double-check the calibration manually using the centroids of strong neutral lines in case a small residual drift is still present. Such sanity check cannot be performed in our data set because the raster is located off-limb and neutral lines are not observed. 

Further, we note that the peak time of the M class flare occurred at about 22:22UT, but the IRIS observation did not detect any signal in the Fe XXI line before the time shown in Figure~\ref{fig:fexxi}. On the other hand, high temperature plasma is already observed before this time in the AIA 131~\AA~channel, showing plasma at 10 MK. This lack of emission in the IRIS Fe XXI line may be caused by the fact that this line is particularly faint, as mentioned above. This also suggests that the non-thermal velocity might have been higher during the early impulsive phase of the flare, before we can observe any Fe XXI emission in IRIS.

\subsection{Model Setup}
\begin{figure*}[h!]
    \centering
    \begin{center}
    \includegraphics[width=0.9\textwidth]{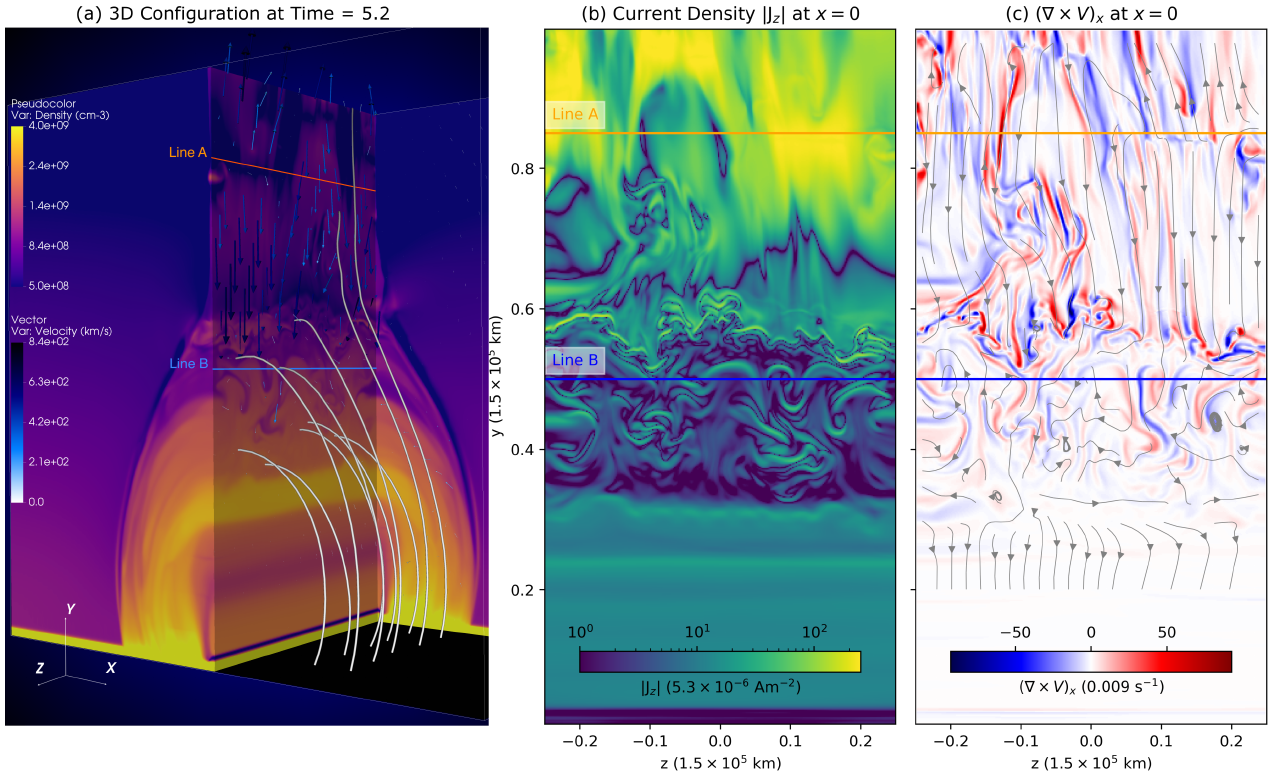}
    \end{center}
    \caption{Distribution of primary variables in a 3D flare model at $time=5.2t_0$. (a) Density and velocity distribution on the chosen planes in 3D. The gray tubes indicate magnetic field lines, and the black/blue arrows are for plasma flows. Along the LOS direction, two sampling lines A and B are chosen on the central plane ($x=0$).
    (b) Current density at the central plane.
    (c) Velocity vorticity component $(\nabla \times V)_x$ on the central plane. The streamlines show velocity fields ($V_y, V_z$).
    %{\color{magenta}{could you choose different values for linesd A and B}? they are not very visible.}
    }
    %\footnote{The displayed streamlings are drawed based on $V_y, V_z$ components on the $x=0$ plane. But plasma flows are 3D in the MHD model}
    \label{fig:mhd_overview}
\end{figure*}

We use the same 3D solar flare model as shown in \cite{Shen2022NatAs...6..317S}, which follows the classic CHSKP configuration of two-ribbon flares. 
In this model, we solve the initial and boundary value problem governed by resistive MHD equations using the public code Athena \citep{Stone2008ApJS..178..137S}.
The system is initialized from a pre-existing vertical Harris-type CS along the y direction in mechanical and thermal equilibrium. Driven by the initial perturbation on magnetic fields \citep[also see][]{Shen2022NatAs...6..317S}, the CS becomes thinner due to the Lorentz-force attraction, and a pair of reconnection jets flow away from the reconnection X-point close to the initial perturbation center. 
The close magnetic loops then appear at the bottom results of the line-tied boundary, in which the magnetic field lines are rooted at the boundary. 
Once the flare loops are well-formed, we start the 3D simulations by symmetrically extending all primary variables from the 2D domain to the third direction ($z-$).
The system then self-consistently evolves, and a set of reconnection-driven flare phenomena are seen, including Alfv\'{e}nic bi-directional reconnection outflows, the termination shocks, and the turbulent interface region below the extended reconnection current sheet and above the flare loops (Figure \ref{fig:mhd_overview}(a)). 
In this work, we set the primary characteristic parameters as follows: length $L_0 = 1.5 \times 10^8$m, magnetic field strength $B_0=0.001$T, density $\rho_0=2.5 \times 10^{14}$m$^{-3}$, temperature $T_0=1.13 \times 10^8$K, velocity $V_0 = 1.366 \times 10^6$m s$^{-1}$, and time $t_0 = 109.8$s.

To compare our models with observations, we calculate synthetic Fe XXI 1354 \AA\ line profiles that are observable by IRIS in high-temperature flare plasma (e.g., $\sim$11MK).
Once we compute the plasma properties (temperature, density, and velocity) on each cell from the 3D MHD simulation, the synthetic emission can be obtained along any chosen LOS (e.g., Lines A and B in Figure \ref{fig:mhd_overview}) by using the formula \citep[e.g.,][]{Guo2017ApJ...846L..12G}:
\begin{equation}
    I_{\nu} = \frac{h\nu}{4\pi} \int f_{\nu} n_e n_H g(T_e)\,dl\,
    \label{eq: emiss_Iv}
\end{equation}
where $\nu$ indicates the frequency of emission lines, and $l$ is the integration path. $n_e, n_H \sim 0.83n_e$ are the electron and proton densities, which are obtained from MHD plasma density in the following calculations. $g(T_e)$ is the contribution function, which can be obtained from the CHIANTI database \citep{Zanna2021ApJ...909...38D}. $f_\nu$ indicates the variation of velocity distribution function due to plasma flows along the LOS, given by
\begin{equation}
    f_\nu = \frac{1}{\pi^{1/2}\Delta \nu} exp(-(\frac{\Delta\nu + \nu_0 v_{l}/c}{\Delta \nu_D})^2).
\end{equation}
Here $v_{l}$ is the plasma flow speed along the LOS, $c$ is the speed of light. $\nu_0$ is the rest frequency, and $\Delta = \nu - \nu_0$ means the offset frequency accordingly. The thermal broadening can be calculated by
\begin{equation}
    \Delta \nu_D = \frac{\nu_0}{c} \sqrt{\frac{2kT}{m}},
\end{equation}
 where $k$ is the Boltzmann constant, $T$ is the temperature, $m$ is the atomic mass of the chosen ion (e.g., Fe in this work).

\begin{figure*}[t!]
    \centering
    \begin{center}
    \includegraphics[width=0.9\textwidth]{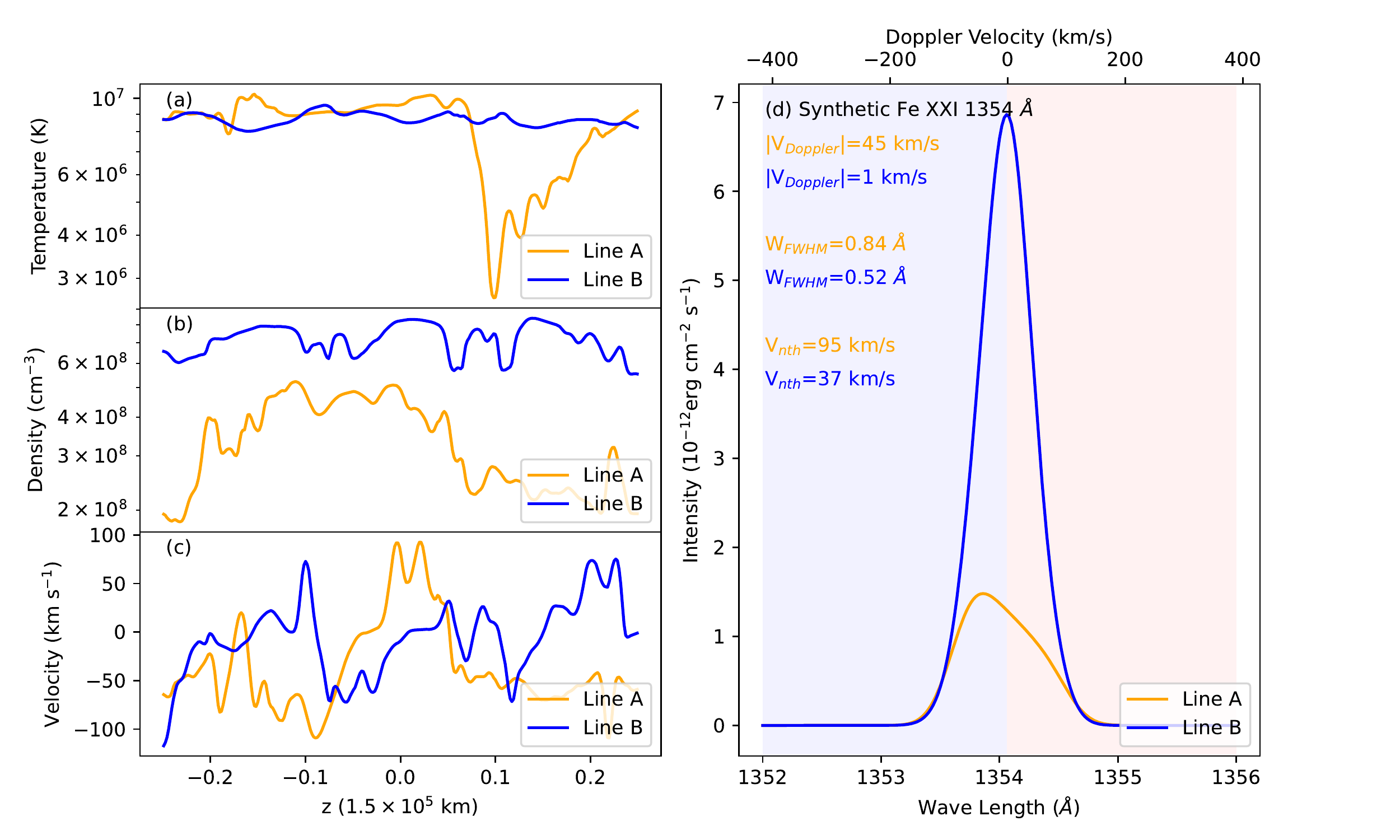}
    \end{center}
    \caption{Synthetic Fe XXI 1354.08 \AA\ emission along the sampling lines A and B shown in Figure \ref{fig:mhd_overview}.
    (a-c) Temperature, density, and velocity profiles along the two sampling lines.
    (d) Synthetic spectra for lines A and B. $v_{Doppler}$ and $v_{nth}$ indicate the Doppler shift velocity and non-thermal velocity, respectively. $W_{FWHM}$ are the line broadenings of each spectral line obtained by measuring the full width at half maximum (FWHM).}
    \label{fig:fig3}
\end{figure*}

\section{Results}
\label{Sect:results}
We calculate the synthetic emission line profiles of the Fe XXI 1354 \AA\ line when the dynamic flows are well-developed in both the CS and flare cusp regions.
Figure \ref{fig:mhd_overview} shows the magnetic field lines and the distribution of primary variables (density and velocity) at the chosen time (time = 5.2$t_0$, with the chosen characteristic timescale $t_0 \sim 110$~s). 
Above $y \sim 0.6 L_0$, the reconnection downward outflows can be clearly seen from the velocity vectors (Figure \ref{fig:mhd_overview}(a)) and streamlines (Figure \ref{fig:mhd_overview}(c)). The turbulent structures in the interface regions below the CS are well illustrated on the current density map, as shown in Figure \ref{fig:mhd_overview}(b). The high shearing and vortex flow, indicated by the large curl of the velocity field ($\nabla \times V$) and randomly rounding streamlines, are easily found around the chosen sampling Lines A and B as well.

\subsection{Emission Line Profiles}
We first analyze the spectral properties along the sampling Lines A and B. As shown in Figure \ref{fig:mhd_overview}, Line A is located at a relatively higher altitude along the CS, which is close to the primary reconnection X-points indicated by the strongest current density ($J_z$, in Figure \ref{fig:mhd_overview}b) and flow stagnation regions (Figure \ref{fig:mhd_overview}c).
On the other hand, Line B is chosen to go through highly turbulent above loop-top regions, where the SADs-like tenuous down flow features are well-developed as reported by \cite{Shen2022NatAs...6..317S}.
At this time (5.2$t_0$), Line A shows slightly higher temperature and lower density than that of Line B overall (Figure \ref{fig:fig3}). Along the LOS, we note that the plasma is  appreciably cooler than the expected Fe XXI 1354\AA\ line temperature (e.g., Log T$\sim 7.05$~K). This effect occurs because the LOS path is outside the high-temperature CS at some positions. % though the cool plasma does not significantly contribute to the total Fe XXI line emission.
Due to the highly turbulent plasma flows, the temperature, density, and velocity along the sampling lines all show various perturbations. 
The domain perturbation velocity ranges from $\sim +100$ to $\sim -100$ km s$^{-1}$ as shown in Figure \ref{fig:fig3}(c).

\begin{figure*}[t!]
    \centering
    \begin{center}
    \includegraphics[width=0.9\textwidth]{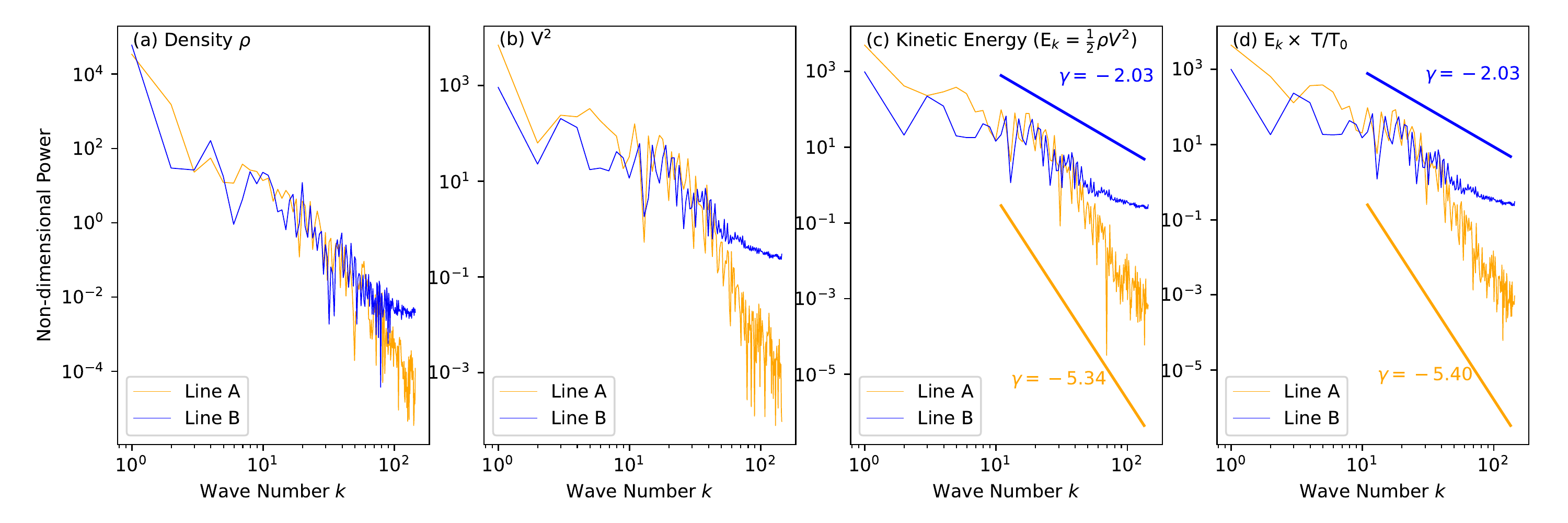}
    \end{center}
    \caption{Fourier power spectrum along the sampling lines A and B for (a) density perturbations, (b) velocity power, (c) kinetic energy perturbations, and (d) temperature-weighted kinetic energy. The peak formation temperature $T_0$ for Fe XXI lines is about 11.5 MK in panel (d). The horizontal axis is wave numbers, and the vertical axis is non-dimensional power. The $\gamma$ indicates the power index by fitting the spectrum profiles with power-law functions ($\sim k^{-\gamma}$).}
    \label{fig:fig3_att}
\end{figure*}

Figure \ref{fig:fig3}(d) shows the synthetic line profiles of the Fe XXI 1354 \AA\ lines for the above two sampling lines. 
We assume that LOS is from $z=-0.25L_0$ to $z=+0.25L_0$ along each sampling line, and the Doppler shift velocity is then plotted based on this geometry accordingly. We note that the LOS direction may also be reversed in the observations, so the red and blue shifted Doppler velocity is a relative value in this plot. 
In this case, Line A shows a larger Doppler shift velocity (about 18 km s$^{-1}$) than that of Line B because the Alfv\'{e}nic downward reconnection outflows around Line A dominate the overall flow behavior. 
We measure the line broadening width using the full width at half maximum (FWHM) of the synthetic emission profiles. Then the non-thermal velocity ($v_{nth}$) can be obtained according to the following equation:
\begin{equation}
    \text{FWHM} = \sqrt{4 \text{ln2} (\frac{\lambda_0}{c}) (\frac{2kT_{eq}}{m} + v_{nth}^2)},
\end{equation}
where the $\lambda_0$ is the central wavelength of Fe XXI 1354 \AA\ line, log$T_{eq}$ = 7.05\ K is the typical formation temperature of this line.
In this particular position, the non-thermal velocity of Lines A ($\sim 95$ km s$^{-1}$) and B ($\sim 37$ km s$^{-1}$) are consistent with the deduced values in recent observations \citep[e.g.,][]{Doschek2014ApJ...788...26D, Warren2018ApJ...854..122W}. However, the synthetic line profile strongly depends on the space and time evolution of the solar flare system, which will be addressed in the next section.
In general, the actual temperature on each pixel (or cell) along the LOS cannot always be exactly at the  $T_{eq}$, in either the observations and MHD models. Therefore, the deduced non-thermal velocity is unavoidably affected by the thermal broadening due to different temperatures, especially for high-temperature plasmas with log$T$ > 7.05\ K.
In the following sections, we will follow this approach (where the equilibrium temperature is usually assumed when calculating the non-thermal broadening) to match the observational data analysis but give a more detailed discussion in Section 3.3.

We analyze the perturbation properties of primary variables along the LOS by using the Fourier transform method to investigate the nature of the broadening of spectral lines. 
In Figure \ref{fig:fig3_att}, we perform the one-dimensional Fourier transform for density ($\rho$), velocity square ($v^2$), and kinetic energy ($E_k$) along the Lines A and B. It is clear that these perturbations appear on all scales (or wave number ($k$) ranges), and the kinetic energy cascades from the large scale to the small scale as well. Furthermore, the spectrum matches the power-law tendency as the prediction in the classic turbulence theories, which could naturally result in the line broadening.
Due to the limitation of the MHD grid sizes, we do not show here the spectrum distribution in high-$k$ ranges where the possible inertial range is expected in turbulence theories.
However, we can estimate quantitatively the turbulence properties in the intermediate ranges by fitting a power-law spectrum ($\sim k^{-\gamma}$) in Figure \ref{fig:fig3_att}(c).
We also check the temperature weighted $E_k$ in according to the chosen temperature (log$T \sim 7.05$ K) for the Fe XXI 1354 \AA\ line. A similar power law distribution pattern can be seen clearly, as shown in Figure \ref{fig:fig3_att} (d). 
We notice that Line A with larger Fe XXI 1354 \AA\ line broadening shows a higher index ($\gamma \sim 5.34$) as compared with the shallower line B ($\gamma \sim 2$). This result indicates that large bulk flows can significantly contribute to the line broadening as shown by the velocity spectrum of Line A, in which the more turbulent flows remain in the low-$k$ (or large scale) ranges.

\begin{figure*}[t!]
    \centering
    \begin{center}
    \includegraphics[width=1.0\textwidth]{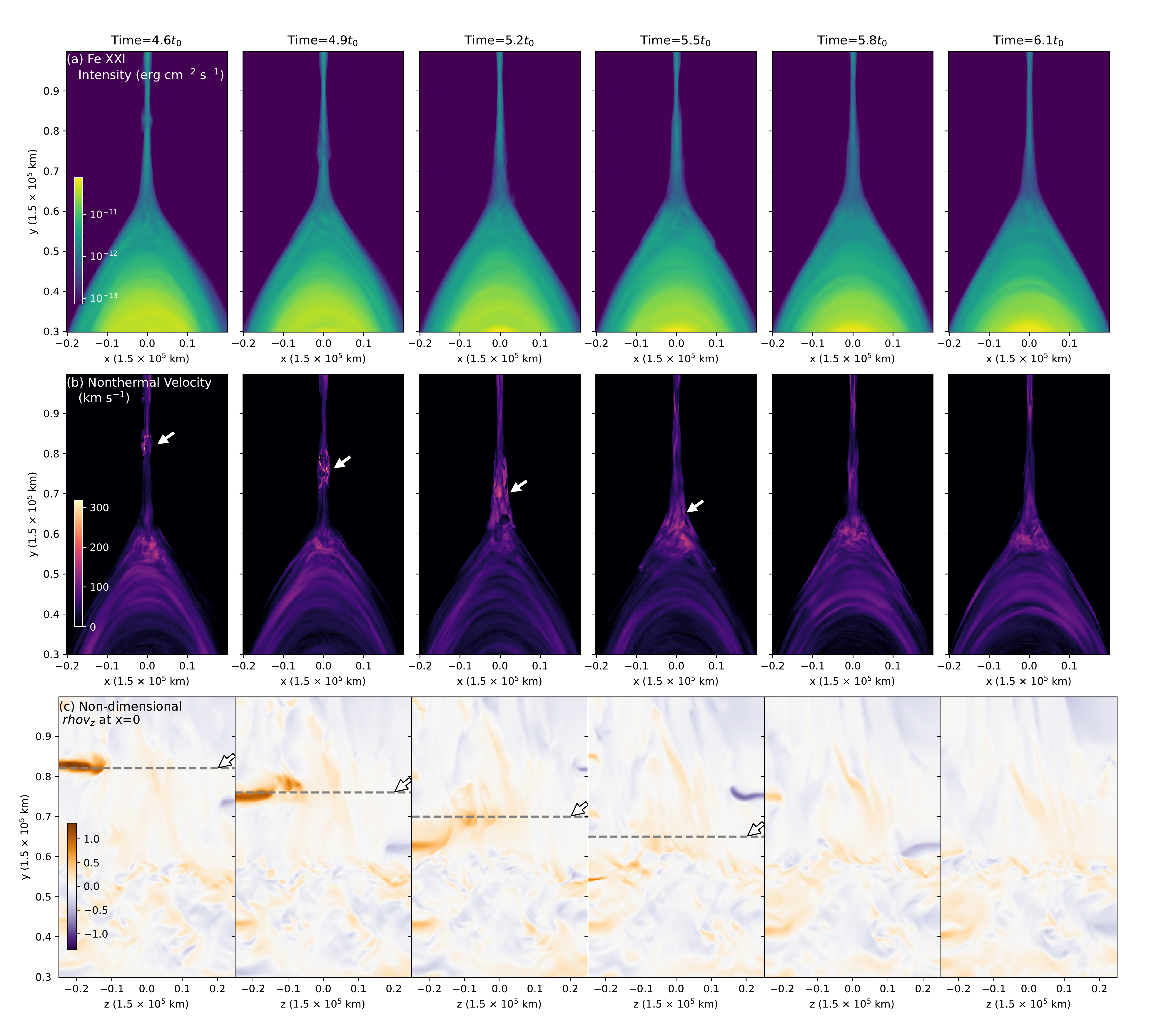}
    \end{center}
    \caption{Synthetic Fe XXI 1354 \AA\ emission maps at different times.
    (a) The peak intensity of 1354 \AA\ spectra line during time $t=4.6t_0$ to $6.1t_0$;
    (b) The deduced non-thermal velocity ($v_{nth}$). The white arrows annotate the downward moving high $v_{nth}$ structures.
    (c) Momentum $\rho V_z$ distribution on the center plane ($zoy$). The dashed lines and arrows indicate the same heights annotated in panel (b).}
    \label{fig:fig4}
\end{figure*}

\subsection{Spatial and Temporal Evolution}

\begin{figure*}[t!]
    \centering
    \begin{center}
    \includegraphics[width=1.0\textwidth]{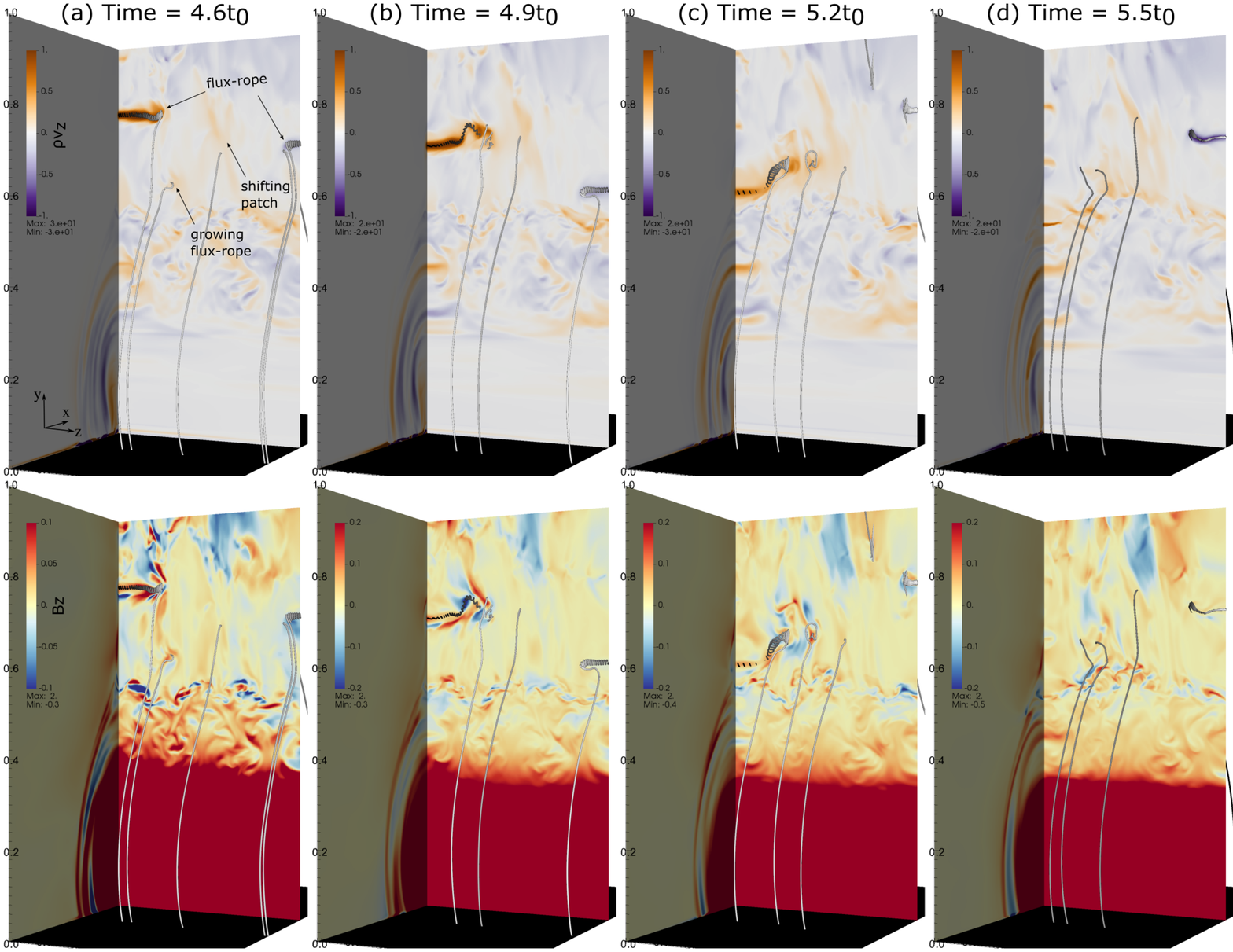}
    \end{center}
    \caption{Distribution of momentum component ($\rho V_z$) and magnetic component $B_z$ on the chosen plane at different times. The upper row is for ($\rho V_z$) in which the gray tubes show magnetic field lines around the well-developed magnetic flux-ropes, the growing magnetic flux-rope, and shift-moving reconnection outflow patches in $z-$ direction.
    The low panels are for $B_z$ with the same magnetic field lines.}
    \label{fig:fig_rhovz_bz_3d}
\end{figure*}

%Show 2D map about line-width and non-thermal velocity
In Figure \ref{fig:fig4}, we plot the distribution and temporal evolution of synthetic Fe XXI 1354 \AA\ emissions with a viewpoint such that the plasma sheet is viewed edge-on.
At each MHD simulation cell on the $x-y$ plane, we obtain the synthetic spectral lines assuming that the LOS is along the $z-$ direction. The peak intensity along with each spectra line is shown in Figure \ref{fig:fig4}(a), in which the bright sheet structure and flare cusp regions can be clearly seen as predicted by the standard solar flare model. The non-thermal velocity is calculated and shown in Figure \ref{fig:fig4}(b).

Figure \ref{fig:fig4} shows that the region where there is strong non-thermal line broadening (up to $\sim 300$ km s$^{-1}$) is dynamically evolving in both the CS and flare cusp regions. 
The region of high non-thermal broadening flows downwards along the CS from the primary reconnection X-point region to the flare cusp region. The white arrows in Figure \ref{fig:fig4}(b) illustrate this process during about one characteristic timescale from $t=4.6t_0$ to $5.5t_0$. 
The downward-moving patch of high non-thermal broadening finally consolidates into the cusp region after $t \sim 5.5t_0$, and causes extended regions of high non-thermal broadening above the flare loop-top.  

In order to understand the magnetic field topology and plasma flow properties in these high-$v_{nth}$ regions, we plot out the horizontal momentum component ($\rho V_z$) in Figure \ref{fig:fig4}(c) and magnetic component ($B_z$) in Figure \ref{fig:fig_rhovz_bz_3d}. Because the emission intensity of the Fe XXI line is proportional to $\rho^2$ and is primarily affected by horizontal flow $V_z$, the plasma momentum ($\rho V_z$) should be a sensitive factor to distinguish the effects of different structures on emission line profiles and line broadening. 
As shown in Figure \ref{fig:fig4}(c), the dashed lines and white arrows highlight  the heights characterized by  enhanced $v_{nth}$, similarly to  Figure \ref{fig:fig4}(b). It is clear that the position of the strong horizontal momentum component coincides with such heights. 
The strongest momentum regions are usually associated with well-developed magnetic flux-ropes. Figure \ref{fig:fig_rhovz_bz_3d} shows a three-dimensional view, with the background color in the first row indicating the $\rho V_z$, same as in Figure \ref{fig:fig4}(c).
At time=4.6$t_0$, two of these flux-ropes are highlighted by plotting the chosen helical magnetic field lines where the strongest  momentum component appears. At  later times (4.9 $\sim$ 5.2$t_0$), these two flux-ropes are associated with strong $\rho V_z$ components that move downwards to lower heights.
In fact, the strong momentum components ($\rho V_z$) are due to the appearance of a guide field $B_z$ in such a turbulent reconnection current sheet. As shown in the second row of Figure \ref{fig:fig_rhovz_bz_3d}, the guide field $B_z$ widely appears with a very turbulent behavior inside the whole current sheet. The typical relative strength of $B_z$ to the total background ($B$) ranges from $\sim$5\% to 10\%, and can be larger than $\sim$30\% at particular positions and times (indicated by the dark-blue color in Figure \ref{fig:fig_rhovz_bz_3d}).
It is then not surprising that the well-developed flux-ropes with strong momentum components ($\rho V_z$) are generally found in high $B_z$ regions.
In addition, other structures all have significant contributions to the momentum component $\rho V_z$, 
including the growing flux-ropes and very turbulent reconnection outflows with remarkable horizontal flow components along the $z-$ direction.
As shown in Figure \ref{fig:fig_rhovz_bz_3d}, well-developed flux-ropes are relatively rare in the whole current sheet region. The fluctuations in $\rho V_z$ can be commonly found around the growing flux-ropes and turbulent reconnection outflow patches as well. 
For example, a large number of growing flux-ropes appear at the center region at time = 5.2$t_0$, which is consistent with the high $v_{nth}$ region above $y = 0.7L_0$. Meanwhile, the nearby fully developed flux rope is located at a much lower altitude ($y \sim 0.65L_0$).

%\begin{figure*}[t!]
%    \centering
%    \begin{center}
%    \includegraphics[width=1.0\textwidth]{./figs/fig7_1d.png}
%    \end{center}
%    \caption{The scatter plot shows non-thermal velocity distribution along the CS direction ($y-$) at the chosen times. The colors indicate the mean amplitude of plasma flow velocity along LOS by computing the non-dimensional $<\delta v_z^2>$. Four arrows annotate the downward moving structures appearing on line-broadening maps, Figure \ref{fig:fig4}(b-c).}
%    \label{fig:fig7_1d}
%\end{figure*}

%In Figure \ref{fig:fig7_1d}, we show the deduced non-thermal velocity at different heights colored by the turbulence strength of plasma velocity perturbations: $V_{turb} \equiv \sqrt{\frac{1}{n_z}\sum{{v_z}_i}^2}/V_A$ to describe the turbulent intensity along the LOS. 
%Here, ${{v_z}_i} \equiv v_z - \Bar{v_z}$ is the turbulent fluctuation of each cell to the mean velocity, $V_A$ is the characteristic Alfv\'{e}n speed, and $n_z$ is the total cell number along the LOS.
%The black arrows highlight the downward moving structures with high line-broadening and large non-thermal velocity (also see white arrows in Figure \ref{fig:fig4}). 
%It can be seen that the downward moving high-broadening patches are indeed the most turbulent structures with the largest $V_{turb}$.
%However, the largest non-thermal velocity is not always cospatial with the greatest velocity perturbation because the Fe XXI emission also strongly depends on temperature and density distributions.

\begin{figure*}[t!]
    \centering
    \begin{center}
    \includegraphics[width=0.7\textwidth]{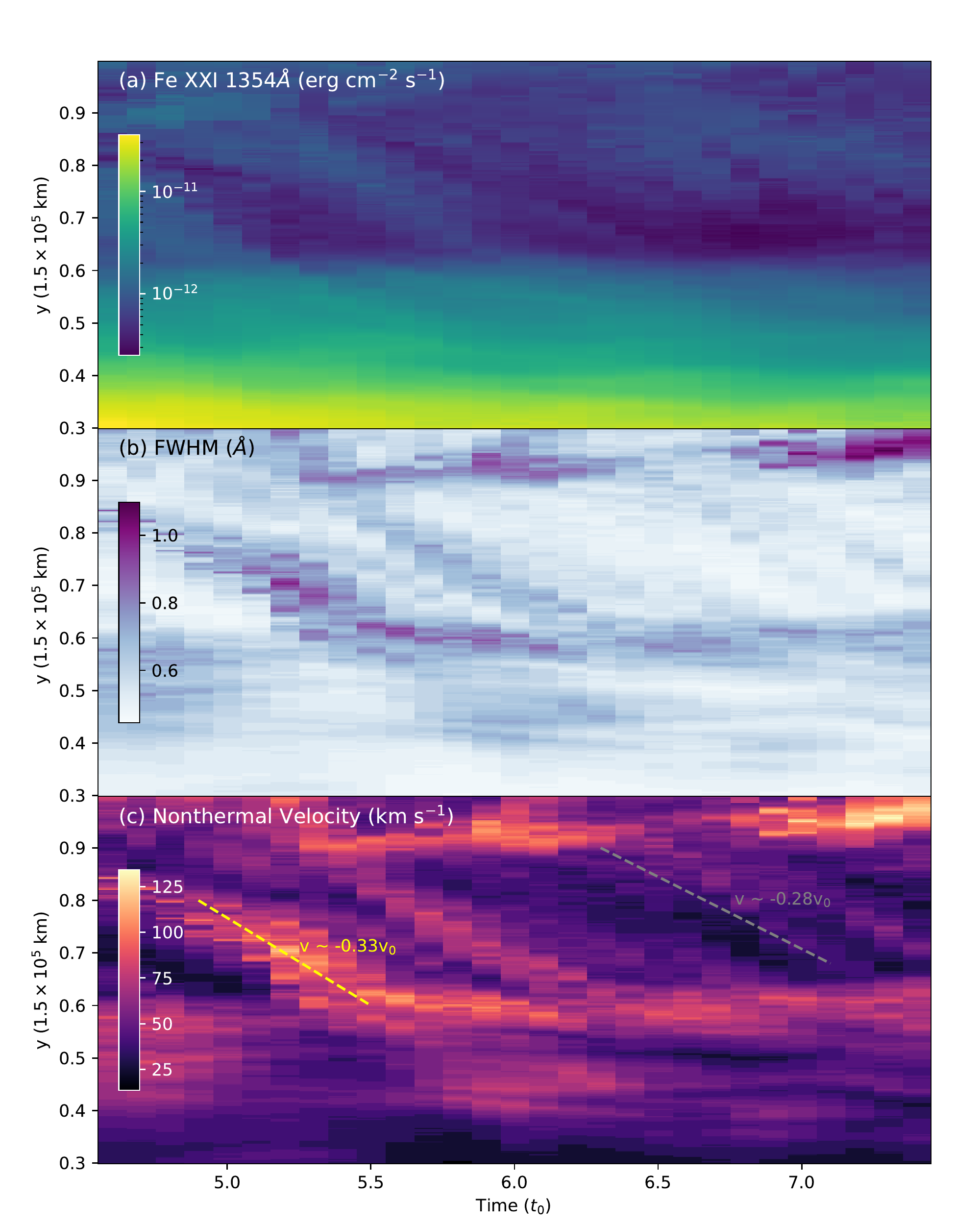}
    \end{center}
    \caption{Time-distance maps along the CS direction at the system center ($x=0$). Panel (a) is for maximum Fe XXI 1354\AA\ intensity, (b)-(c) are for FWHM and the non-thermal velocity ($v_{nth}$) as functions of time. At each height, the median values over all non-zero values are shown here. The red and gray dashed lines in panel (c) indicate the downwards moving features with the speed of $\sim$0.33 and $\sim$0.28$v_0$, respectively.}
    \label{fig:fig6}
\end{figure*}

The downward high-$v_{nth}$ features commonly exist for a longer duration in the reconnection process.
Figure \ref{fig:fig6} displays the time–distance map (or “stack plot”), showing the distribution of the Fe XXI 1354 \AA\, FWHM, and non-thermal velocity along at the CS center ($x = 0$) as a function of time. 
The high-$v_{nth}$ emission flows away from the primary X-point regions nearby $y \sim 0.9$, as can be clearly seen in Figure \ref{fig:fig6}(b) and (c).
These high-$v_{nth}$ patches are most like to originate from the reconnection X-point sites (around $y \sim 0.9 L_0$), and spread with the reconnection outflows to the lower end of the CS.
At different altitudes along the CS, there are two obvious regions of high nonthermal broadening: one is close to the primary reconnection X-point site and another is located above the flare loop-top. 
Above the flare loop-top region, the non-thermal velocity is strong but highly depends on the more complex plasma dynamics in this interface region. It could be roughly the same or decrease with height, as shown in Figure \ref{fig:fig6}(c).
We also notice that high nonthermal-broadenings also appear in upward flowing plasma near the upper boundary ($y \sim 1.0$).

\subsection{Distribution of Non-thermal Velocity}
\begin{figure*}[t!]
    \centering
    \begin{center}
    \includegraphics[width=0.8\textwidth]{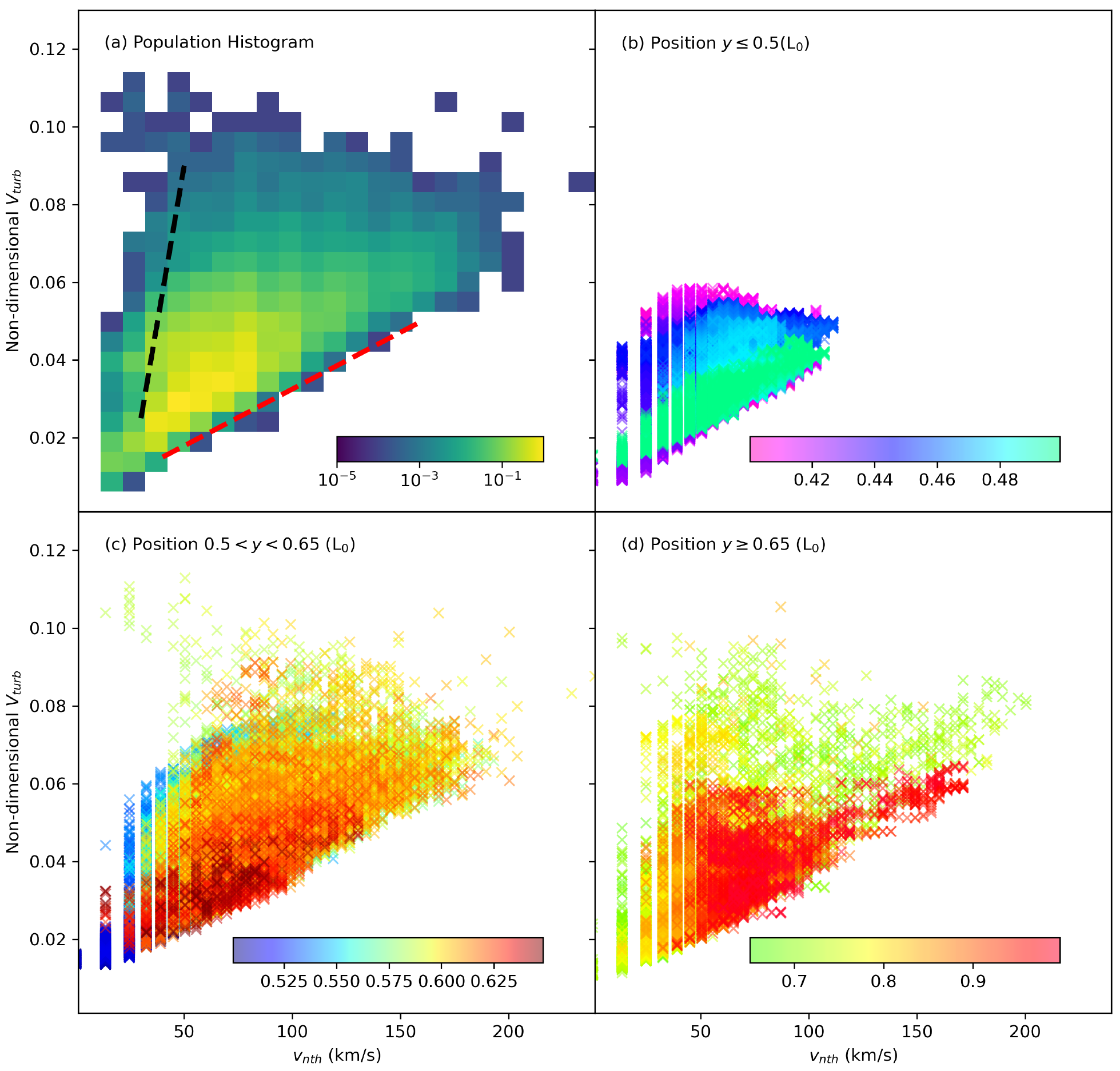}
    \end{center}
    \caption{Distribution of plasma turbulence strength $V_{turb}$ versus non-thermal velocity $v_{nth}$. 
    Panel (a) shows the population of samples in the 2D histogram of $v_{nth}$-$V_{turb}$ map. 
    The red and black dashed lines indicate two typical monotone correlations between $v_{nth}$ and $V_{turb}$.
    Panels (b)(c)(d) show each sample on the $v_{nth}$-$V_{turb}$ maps at different heights: near the flare loop-top region ($y \leq 0.5L_0$), the above-the-looptop region ($0.5L_0 < y < 0.65L_0$), and CS regions ($y \geq 0.65 L_0$), respectively.}
    \label{fig:fig_vnth_vs_vzpert}
\end{figure*}
It is interesting to examine the correlation between the non-thermal velocity and plasma turbulence over a sample of locations in both the CS and flare loop-top regions during a relatively long period from 4.6 $\sim$ 7.4 $t_0$.
We introduce a non-dimensional parameter $V_{turb} \equiv \sqrt{\frac{1}{n_z}\sum{{v_z}_i}^2}/V_A$ to describe the turbulence strength of plasma flows along the LOS. Here, ${{v_z}_i} \equiv v_z - \Bar{v_z}$ is the turbulent fluctuation of each cell to the mean velocity $\Bar{v_z}$, $V_A$ is the characteristic Alfv\'{e}n speed, and $n_z$ is the total cell number along the LOS.
Figure \ref{fig:fig_vnth_vs_vzpert} plots the two-dimensional population histogram of $V_{turb}$ and non-thermal velocity ($v_{nth}$). In Figure \ref{fig:fig_vnth_vs_vzpert}(a), the background colors indicate the population of samples on the $v_{nth}$ - $V_{turb}$ map. An overarching feature of these plots is that the $v_{nth}$ distribution is clearly proportional to the turbulence strength ($V_{turb}$) because stronger plasma turbulent flows can naturally cause larger line-broadening, as discussed in the previous sections. 
However, the detailed distribution is more widely distributed with the increase of $V_{turb}$, and will not likely be fitted by using one simple linear line. Here, we annotate one typical growth direction using a red dashed line and a deviation direction using a black dashed line in Figure \ref{fig:fig_vnth_vs_vzpert}(a), where the most dominant samples appear. The samples around the red line show higher non-thermal velocities with comparable turbulence strength compared with those around the black line.

The deviation among the black and red dashed lines in Figure  \ref{fig:fig_vnth_vs_vzpert}(a) could be due to the variation of plasma flows and turbulence properties at different heights, especially in the CS region and flare cusp regions where the plasma density and plasma $\beta$ environments are largely different. Therefore, we display all samples on the $v_{nth}$-$V_{turb}$ map with their height information, in which colors indicate $y-$ position of each sample in Figure \ref{fig:fig_vnth_vs_vzpert}(b-d).
We separate all samples into three groups ($y \leq 0.5L_0$, $0.5L_0 < y < 0.65L_0$, and $y \geq 0.65 L_0$) to make the difference more visible. 

Figure \ref{fig:fig_vnth_vs_vzpert}(b) shows the first group of samples with the lowest heights below $y \sim 0.5L_0$, in which both the $v_{nth}$ and $V_{turb}$ are small. Because this group of samples is close to the more dense flare loop-top regions, where the turbulent flows are expected to be relatively lower than those in  the above-the-looptop regions, $v_{nth}$ is naturally small (e.g., $< \sim 100$\ km/s).

The samples in the second group (Figure \ref{fig:fig_vnth_vs_vzpert}(c)) in the very turbulent flare cusp regions have stronger turbulence  and larger non-thermal velocities.
Similar high-$v_{nth}$ and high-$V_{turb}$ can be seen in the CS region as shown in Figure \ref{fig:fig_vnth_vs_vzpert}(d) as well.
Samples in both the CS and the above loop-top regions tend to follow the growth direction marked by the red dashed line in Figure \ref{fig:fig_vnth_vs_vzpert}(a), except small abnormal patches with low $v_{nth}$ and high $V_{turb}$ that usually appear in the low end of the CS ($y < \sim 0.8L_0$) and upper the loop top region ($y > \sim 0.57L_0$). In fact, these abnormal samples that have departed from the red line direction may indicate strong bulk flows which will be discussed in the following section.

It is worth noticing that thermal broadening also contributes to the emission line broadening and has an impact on the corresponding $v_{nth}$ distribution, especially when the plasma is hotter than the equilibrium temperature of Fe XXI 1354\AA\ at log$T_{eq}$ $\sim$ 7.05\ K.
In the above investigations, there is a set of samples in our model that are slightly lower in temperature than the Fe XXI 1354\AA\ temperature ($T_{eq}$) (e.g., see Figure \ref{fig:fig3}(a)).
Therefore, we investigate the $v_{nth}$ distribution in different temperature environments. We maintain the plasma density the same as in the above analysis but scale the temperature to match the observations (e.g., log$T$ 7.0 $\sim$ 7.2 K shown in Figure \ref{fig:xrt_teem}).
In the non-dimensional MHD models, the characteristic temperature can be expressed by $T_0 = {{B_0}^2 \beta_0}/{(2 \mu \rho_0 \tilde{R}})$ and Alfv\'{e}n velocity is $V_A = \sqrt{{B_0}^2/(\mu \rho_0)}$. 
Here, $\beta_0$, $\tilde{R}$, and $\mu$ are background plasma $\beta$, gas constant, and magnetic permeability, respectively.
Thus, the scaling of temperature ($T_0$) with corresponding changes in characteristic velocity ($V_A$) allows the MHD model to be scalable by maintaining the plasma $\beta_0$ as the same. 
Therefore, we increased the simulated temperature to a factor of $1.4 \times$ higher than that in the original MHD model result, and the velocity is also increased by a factor of $\sqrt{1.4}$ accordingly.
As a result, the samples in the model can cover a higher temperature range from log$T$ $\sim 6.9$ to $\sim$7.2K.

\begin{figure*}[t!]
    \centering
    \begin{center}
    \includegraphics[width=0.8\textwidth]{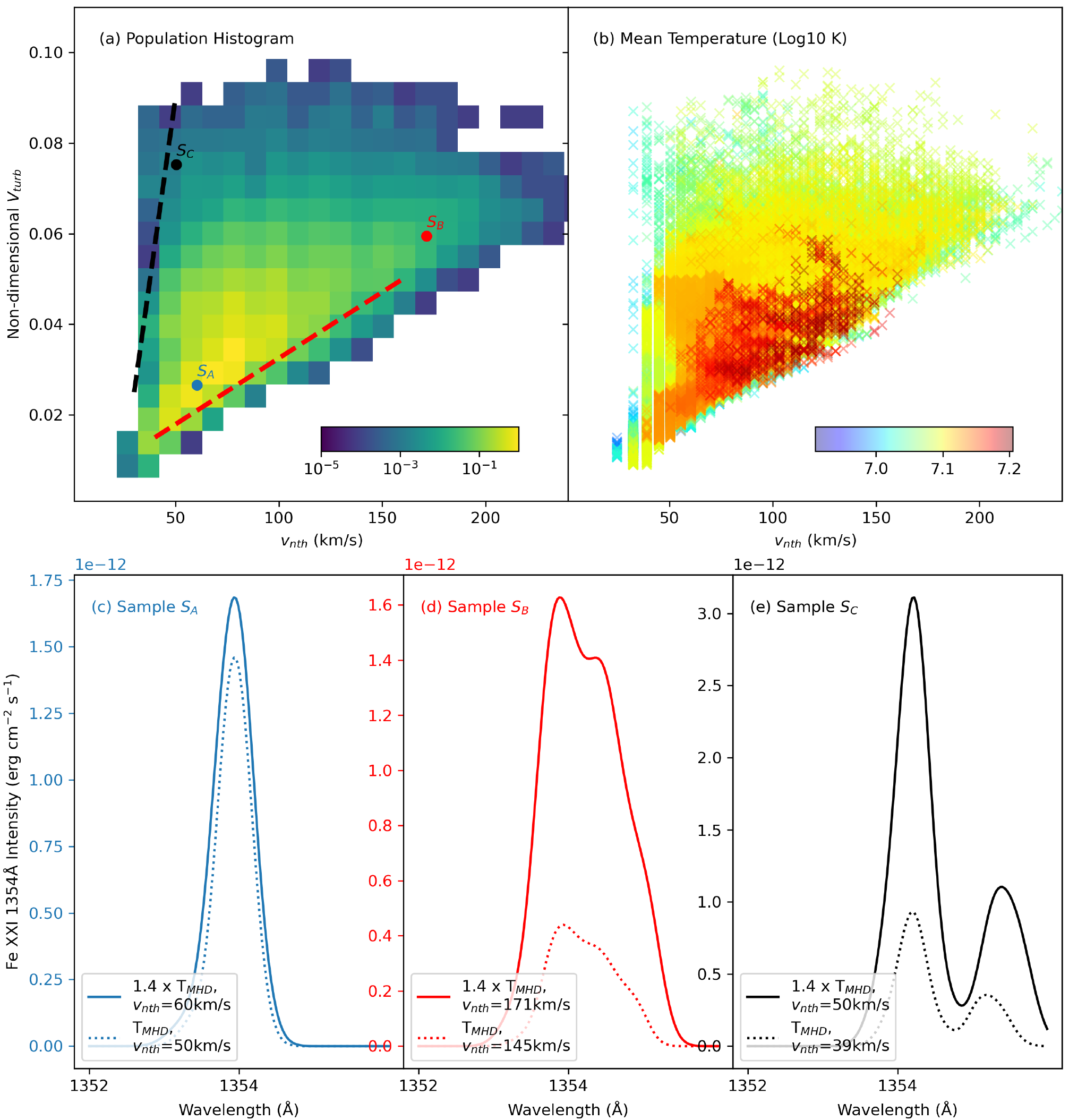}
    \end{center}
    \caption{Distribution of the deduced non-thermal velocity $v_{nth}$ in the scaled temperature case. 
    Panels (a)(b) show $v_{nth}$-$V_{turb}$ map in a $1.4 \times$ higher temperature situation compared with the original temperature ($T_{MHD}$).
    The red and black dashed lines are the same as in Figure \ref{fig:fig_vnth_vs_vzpert}(a).
    Panels (c-e) show Fe XXI 1354\AA\ line profiles at the chosen points: $S_A$, $S_B$, and $S_C$ marked in panel (a). The solid lines are for $1.4 \times$ scaled temperature cases and the dotted lines are for $T_{MHD}$ cases.}
    \label{fig:fig_vnth_ditri_te}
\end{figure*}

Figure \ref{fig:fig_vnth_ditri_te}(a) shows the population histogram of $V_{turb}$ and $v_{nth}$ in the scaled temperature case. The red and black dashed lines are exactly the same as in Figure \ref{fig:fig_vnth_vs_vzpert}(a) as well. 
The distribution feature of $v_{nth}$ is consistent with the above results in the original MHD temperature case ($T_{MHD}$), though the absolute values of $v_{nth}$ are slightly larger by about 10\% $\sim$ 20\%. 
These larger $v_{nth}$ are partially caused by the scaling process of MHD models, because the characteristic velocity increased by about $\sim 18$\%, which could lead to larger Doppler shift velocities and corresponding larger line broadenings.
The mean temperature of each sample is displayed by different colors in Figure \ref{fig:fig_vnth_ditri_te}(b).
%Because the characteristic velocity is already to $\sim 1.18 \times$ higher, the increasing of absolute values of $v_{nth}$ due to thermal broadenings is indeed diminutive.
It is interesting to note that the highest temperature (log$T \sim 7.2$K) does not necessarily indicate the largest $v_{nth}$ if the turbulence strength remains low (the dark-red samples), which suggests that the plasma perturbation behaviors have more crucial impacts on the non-thermal velocity distribution than the thermal broadening itself.

Figure \ref{fig:fig_vnth_ditri_te}(c-e) shows Fe XXI 1354\AA\ line profiles in three typical regions: $S_A$, $S_B$, and $S_C$. 
Sample $S_A$ represents the most dominant features with low turbulence strength and low non-thermal velocity. The line profile basically matches the Gaussian-type shape with minor line broadening due to the relatively weak turbulent flows. 
Sample $S_B$ is at the high $v_{nth}$ end with strong $V_{turb}$, where the strong plasma perturbations contribute to a remarkable line-broadening in both the $1.4 \times$ higher and slightly lower $T_{MHD}$ cases. In addition, the plasma bulk flows also cause the second emission peak at $\sim 1354.4$\AA\, which makes the line profile depart from a Gaussian distribution and causes a wider $v_{nth}$. 
Sample $S_C$ indicates the abnormally high $V_{turb}$ with low non-thermal velocity. The reason can be found in the line profiles which show two separate emission peaks: one is for the dominant Fe XXI 1354\AA\ with a smaller Doppler velocity and the other one is centered at $\sim 1355.4$\AA\ due to the enormous plasma bulk flow, such as the well-developed flux-ropes as shown in Figure \ref{fig:fig_rhovz_bz_3d}. 
In general, the turbulence strength ($V_{turb}$) only refers to mean perturbation features, and it could be too simple to represent complex flow properties including randomly turbulent flows and large bulk flows.
Therefore, the appearance of these abnormal samples around the black dashed line in Figure \ref{fig:fig_vnth_ditri_te} suggests that the investigation of plasma turbulence based on the emission line broadening features must consider the different fine structures in the CS (e.g., plasma blobs or flux-ropes) and flare loop-top regions (e.g., macro plasma instabilities or SADs in \cite{Shen2022NatAs...6..317S}).

\subsection{Non-thermal Velocity Across the CS}

\begin{figure*}[t!]
    \centering
    \begin{center}
    \includegraphics[width=1.0\textwidth]{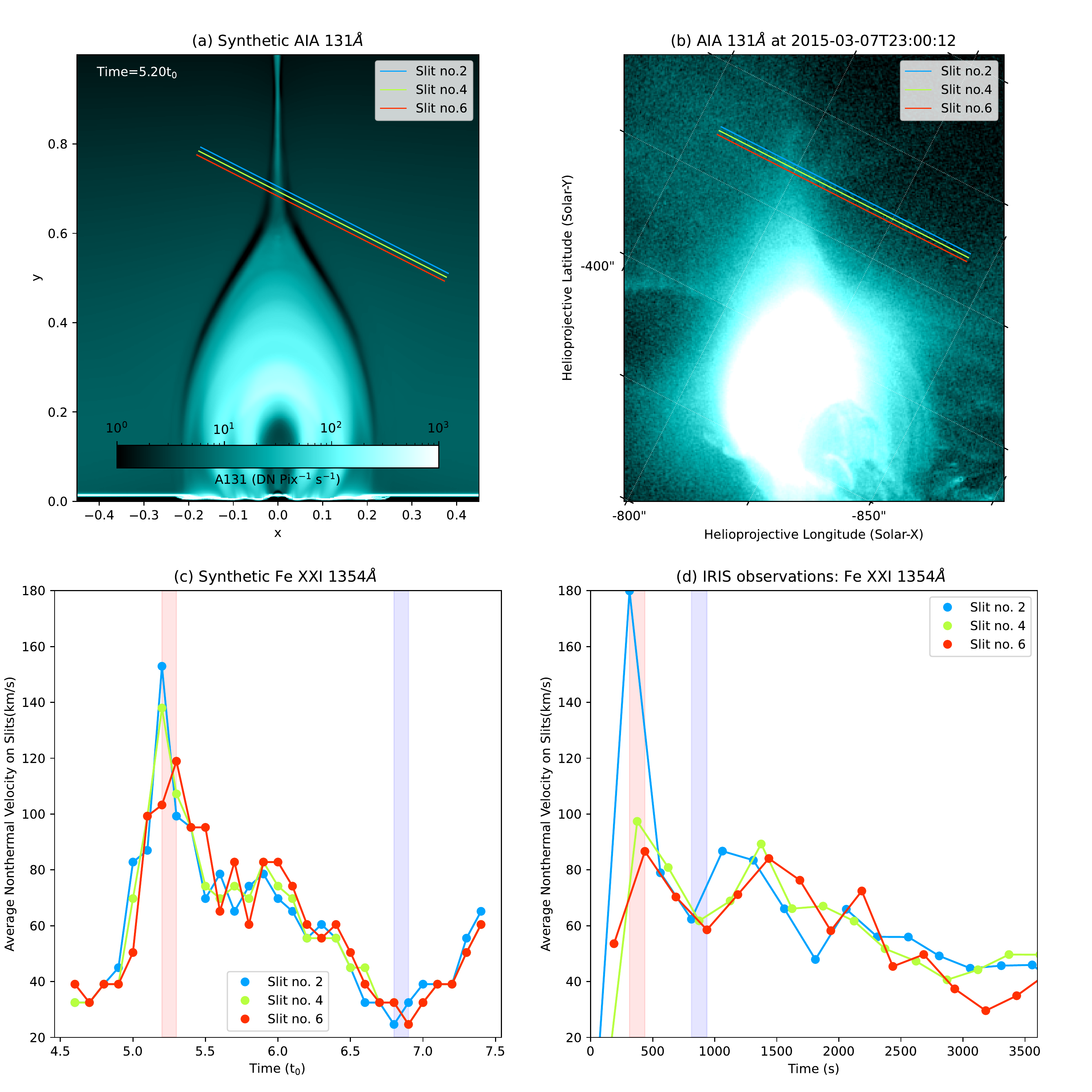}
    \end{center}
    \caption{Modeled and observed non-thermal velocity of IRIS Fe XXI line. The left panels show: (a) the synthetic SDO/AIA 131 image, and (c) the non-thermal velocity calculated by using the averaged Fe XXI 1354 \AA\ profiles along each IRIS slit. The colored solid lines in panels (a) and (c) indicate the IRIS slits with the length of $\sim 9.3 \times 10^4$ km (or $\sim$ 128 arcsec). 
    The right panels (b) and (d) are for SDO/AIA 131 observations at 2015-03-07T23:00:12 and IRIS raster results, respectively. The start time on the horizontal axis in panel (d) is 2015-03-07T22:55:51.
    The red and blue shadowed regions  display the evolution  of the non-thermal velocity on different slits.  
    }
    \label{fig:fig9}
\end{figure*}

In this section, we investigate the averaged Fe XXI lines profiles across the CS and compare them to the IRIS observations during the 2015-03-07 flare.
We set the position of the simulated IRIS slits above the flare loop top region in the MHD model as shown by Figure \ref{fig:fig9}(a). The synthetic SDO/AIA 131 image in this plot is obtained by assuming that the LOS is along the $z-$ direction. The synthetic bright flare loops, cusp region, and extending CS regions all match well the SDO/AIA images of the solar flare under study (Figure \ref{fig:fig9}a) well.
It is clear that the IRIS slits are located just above the bright flare arches in AIA 131 during this eruption event (also see Figure \ref{fig:obs_overview}). Therefore, we chose a height of $y \sim 0.62$L$_0$ in the MHD model to simulate the IRIS raster in the following analysis.
As the reconnected magnetic flux is accumulated at the solar surface, the flare loop system gradually grows in both vertical and horizontal directions. Thus, the relative position of the IRIS slits to the flare cusp region gradually changes as a function of time. 
Because our MHD model is focused on the fast magnetic reconnection process in solar flares, it is reasonable to compare the predicted SDO/AIA emissions with the observed flare just after the impulsive phase when the post-flare loops are well-formed but the bursty magnetic energy release is still on-going.
To simplify the analysis, we show a particular SDO/AIA observation at the starting time of IRIS observations (2015-03-07T23:00) in Figure \ref{fig:fig9}(b), and keep the simulated IRIS slits in the same position in this work.

Along each slit in Figure \ref{fig:fig9}(a), we average the Fe XXI 1394\AA\ line profiles over all MHD cells with $I > \frac{I_{peak}}{e}$. Here $I$ is the maximum intensity of the Fe XXI line at each cell, and $I_{peak}$ indicates the peak $I$ across the CS on this slit. The non-thermal velocity of each slit is then calculated using the averaged line profile, assuming log$T$ = 7.05K for the strongest Fe XXI emission. 
Figure \ref{fig:fig9}(c) shows the non-thermal velocity variation with time on the chosen slits, No. 2, 4, and 6.
An overall feature is that the $v_{nth}$ variation tendency on these three slits is consistent with each other. For example, the local maximum $v_{nth}$ peak (red shadowed region) appears initially at times 5.2 on slit 2 and then at 5.4$t_0$ for slit 6. A similar tendency is also clear around time 6.8$t_0$, as highlighted by the blue shadowed region. 
By comparing the $v_{nth}$ trends above with Figure \ref{fig:fig6}, one can see that the first  $v_{nth}$ peak (around times 5.3$t_0$) is consistent with downwards moving structures with enhanced Fe XXI line widths between $t=4.6$ to $t=5.5t_0$, as indicated by the yellow dashed line. In the later times ($t=5.5$ to $t=7.4t_0$), the relatively weaker downward structures also caused the second local $v_{nth}$ peak at 5.9$t_0$ and possibly the third peak at 7.4$t_0$ on Figure \ref{fig:fig9}(c).

In the ideal situation when the non-thermal velocity variation is due to the passing downward structures, the characteristic size of detectable downward patches can be estimated as follows:
\begin{equation}
    \delta l + \frac{d_{slit}}{cos(\theta)} \sim v_d \times \delta t.
\end{equation}
Here $\delta l$ is the length of moving patches, $d_{slit}$ is the interval of the IRIS slits, $\theta$ is the intersection angle between slits and the solar surface, $v_d$ is the moving speed of patches, and $\delta t$ is the exposure time of each slit (or interval time between two slits).
For instance, $\delta t \sim 30$\ s for the IRIS Fe XXI observations, $d_{slit}$ is about 726 km, and $\theta \sim 30^{\circ}$. Assuming the downward moving structures are in sub-Alfv\'{e}nic speed (e.g., 300 km/s), a moving enhanced $v_{nth}$ patch with the size of $\delta l \geq \sim 8000$ km should be well recognized. The above estimated $\delta l$ equals $\sim 0.05L_0$, which is reasonable and matches the modeling structures as shown in Figure \ref{fig:fig6}. 

Similar to the synthetic results in MHD models, we also see very similar $v_{nth}$ evolution trends in the IRIS observations (Figure  \ref{fig:fig9}d). The red and blue shadowed regions show one example where the local maximum and minimum $v_{nth}$ has been recognized first on slit 2, then on the following slits 4, and 6.
Remarkably, the evolution of $v_{nth}$ of the  Fe XXI line is comparable between MHD model predictions and IRIS observations, which range from $\sim 20$\ km/s to $180$ km/s during this particular period.
We note that the time-resolution of the deduced $v_{nth}$ variation from the IRIS is limited by the relatively long exposure time ($\sim 30$s). Therefore, Figure \ref{fig:fig9} (d) may unavoidably smooth out fine structures and cannot display the short-period perturbations that are visible in the synthetic modeling results.

\begin{figure*}[t!]
    \centering
    \begin{center}
    \includegraphics[width=0.9\textwidth]{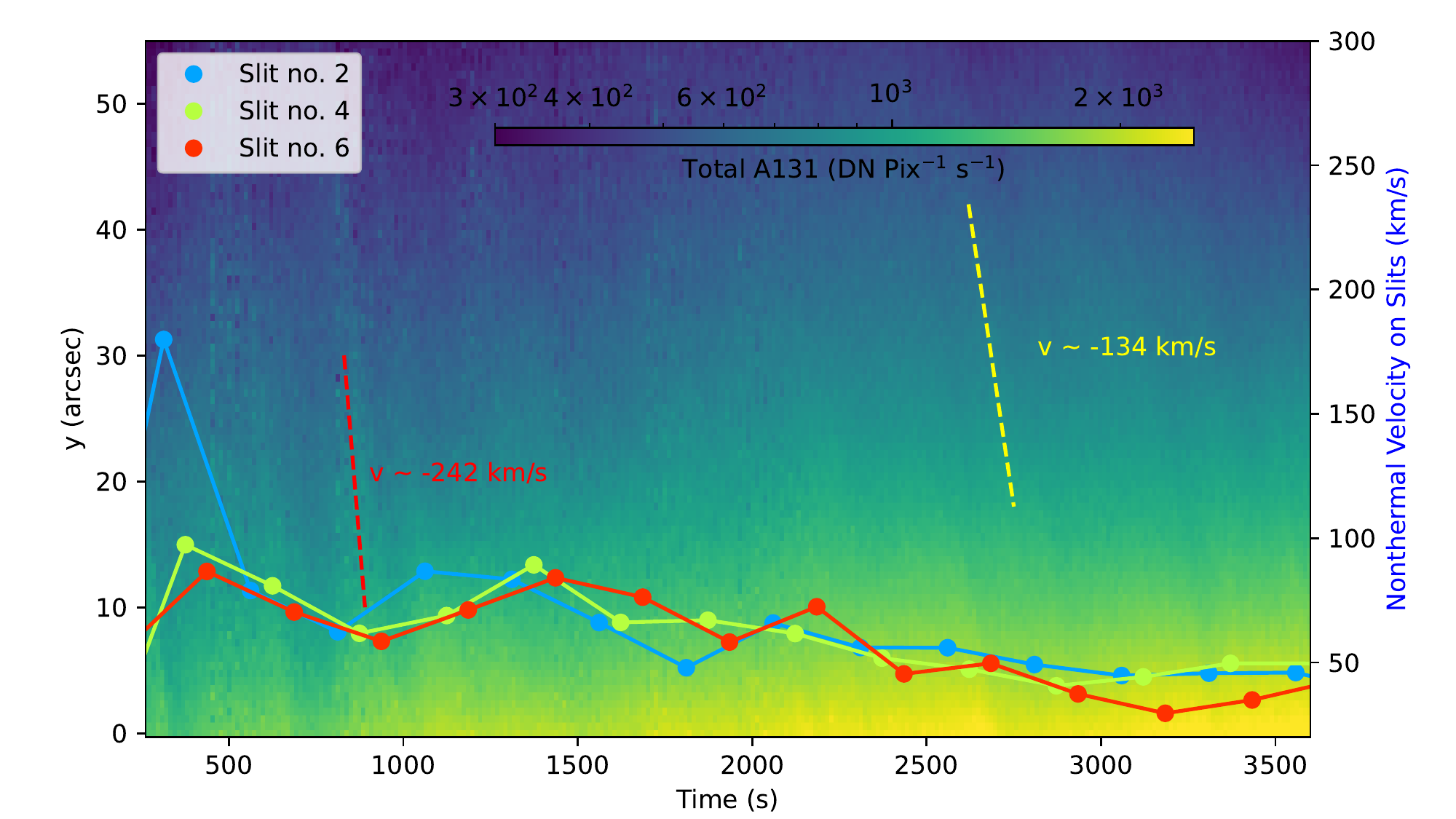}
    \end{center}
    \caption{Time-distance map of SDO/AIA 131\AA\ images in the reconnection region. The total AIA 131\AA\ intensity is calculated by counting 16 pixels at each height around the brightest region above the flare loop top.
    The two dashed lines indicate the typical downwards-moving features with velocities of  $\sim 242$ km/s and $\sim 134$ km/s. 
    The right y-axes are for the $v_{nth}$ of the Fe XXI 1354 \AA\ spectra (solid-dots lines) from the IRIS observations in the same range as that of Figure \ref{fig:fig9}(d). 
    }
    \label{fig:fig10}
\end{figure*}

Due to the similar variations in the MHD predictions and the IRIS observations described above, it is likely that the $v_{nth}$ variation on IRIS Fe XXI lines is due to intermittent downwards moving structures along with the reconnection outflows as shown in the above models. In fact, we find very similar intermittent downwards-moving signals on SDO/AIA 131\AA\ images. 
Figure \ref{fig:fig10} shows the time-distance map of AIA 131\AA\ intensity above the flare loop top regions. 
In this plot, we count all emissions over 16 pixels around the CS region (the bright plasma sheet on AIA 131\AA\ maps) at each height to increase the signal-to-noise ratio. 
Two typical downward features are fitted by using red and yellow dashed lines. These downflow features range from $\sim$ 250 km/s to $\sim$ 100 km/s, which is consistent with the model-predicted moving speed of turbulent structures as shown in Figure \ref{fig:fig6}. 
We also overlay the non-thermal velocity trends on the AIA 131\AA\ stack map as shown by the solid-dots lines. There is no clear correlation between the brightest AIA features and the highest $v_{nth}$ because the AIA intensity is mainly dominated by plasma density in the CS region, while the $v_{nth}$ values are more affected by the turbulent plasma flows.

\section{Discussion and Conclusion}
\label{Sect:discussion}
Using our three-dimensional MHD model based on the classic solar flare configuration, we calculate the synthetic emission profiles of the Fe XXI 1354 \AA\ line observed by the Interface Region Imaging Spectrograph (IRIS) spacecraft. The synthetic line broadening and non-thermal velocities ($v_{nth}$) are obtained around the whole magnetic reconnection regions including the current sheet (CS) and flare loop-top regions. We compare the predicted non-thermal velocities due to the turbulent flows with the IRIS observations during the 2015-03-07 flare.

The main results are summarized as follows:

1. Our MHD model reveals highly turbulent plasma flows both in the CS where the magnetic reconnection sites are located, and in the flare loop top region, an interface region below the CS where the reconnection outflows mix into the flare arcade. Fourier transform analysis of plasma density, velocity, and kinetic energy show that the plasma perturbation spectrum along the LOS ($z-$ direction in this model) follows the power-law tendency of the classic turbulence scenarios.

2. Using the modeled plasma density, temperature, and velocity distribution in 3D, we calculate the emission line profiles in CS and flare loop-top regions in an edge viewing. We found that the non-thermal broadening of Fe XXI 1354\AA\ is mainly due to the plasma bulk flows in high turbulence states. The dominant non-thermal velocity ranges from $\sim$20 km/s to $\sim$180 km/s, and could be up to $\sim$300 km/s in some particular position and times.

3. We obtain a two-dimensional synthetic non-thermal velocity ($v_{nth}$) map and investigate the spatial and temporal evolution features. We find that $v_{nth}$ is dynamically evolving in both the CS and flare loop-top regions. The high $v_{nth}$ structures flow down with a typical speed of $\sim 0.3v_A$ along the CS from the primary reconnection X-point sites to the flare cusp.
By revealing the 3D magnetic field configuration and studying the plasma perturbation amplitude for velocity and density, we confirm that these downwards high $v_{nth}$ structures are due to highly turbulent plasma which is usually associated with complex fine structures inside the CS, such as the growing/well-developed magnetic flux ropes and shifting reconnection outflows with in high guide field ($B_z$) region.

4. We study synthetic $v_{nth}$ distribution versus plasma turbulence strength $V_{trub}$ during a relatively long period. The two-dimensional population histogram map shows that $v_{nth}$ is basically proportional to $V_{trub}$, but with a set of abnormal points due to the large plasma bulk flows. This result indicates that the investigation of plasma turbulence properties based on the emission line broadening features must consider the different fine structures in the CS and flare loop-top regions.

5. We investigate the Fe XXI lines profiles across the CS and make a detailed comparison with the IRIS observations.
The deduced $v_{nth}$ obtained by averaging the Fe XXI emission over the IRIS slit and exposure time ranges from  $\sim$20 km/s to $\sim$160 km/s, which is consistent with the IRIS observations.
Consistent $v_{nth}$ features on different IRIS slits were found in both the synthetic results and the IRIS observations. By comparing the down-flowing high $v_{nth}$ structures and SDO/AIA observations, our results suggest that these downwards-moving fine structures inside the CS could be identified by IRIS due to the variation of the line width of the high-temperature Fe XXI 1354\AA\ line.

In the analysis above, 
%we only set the LOS direction along one specific direction ($z-$ direction) to emphasize the main features of Fe XXI 1354\AA\ line broadening. 
the integration depth along the LOS is limited to the simulation box scale, which is surely shorter than the realistic CME/flare scale. 
Thus, the synthetic Doppler shift of Fe XXI 1354\AA\ line can be significantly affected by several locally large momentum components (e.g., the well-developed flux ropes), and the model-deduced Doppler velocity could be larger than the IRIS observation (Figure \ref{fig:fexxi}(b)). 
Therefore, large-scale solar eruption models are required to make a detailed comparison with the observable Doppler velocity. 

It is worth mentioning that the magnetic configuration of our MHD model may be largely different from the 2015-03-07 flare event. Therefore, we can not exactly compare the model synthetic results with IRIS observations over a very long interval (e.g., $\sim$ 4 hours observation as shown in Figure \ref{fig:fexxi}).
However, the down-flowing high $v_{nth}$ structures can be expected to exist in the model commonly (also see Figure \ref{fig:fig6}). Thus, it is reasonable that to choose  one short period evolution from the model (4.6 $\sim$ 7.4 $t_0$) during the magnetic reconnection process to compare with the IRIS observations.
The speed of down-flowing high $v_{nth}$ structures are predicted to be less than one-third of the Afv\'{e}nic speed in the reconnection region \ref{fig:fig6}), which matches the downflow features observed by SDO/AIA if we assume that  the local background Alfv\'{e}n speed is reduced by a factor of two as compared with the characteristic speed in the model.
Due to the lack of accurate magnetic field information around the CS, further accurate studies on $v_{nth}$ variation require more effort on both the MHD modeling and observational side. In particular, the Multi-Slit Explorer (MUSE), which will be launched in 2027, will be ideal for observing the current sheet and above the loop-top region given its multi-slit configuration and will provide groundbreaking observations of this region as a whole.

%\section{Additional Requirements}
%For additional requirements for specific article types and further information please refer to \href{http://www.frontiersin.org/about/AuthorGuidelines#AdditionalRequirements}{Author Guidelines}.

\section*{Conflict of Interest Statement}
%All financial, commercial or other relationships that might be perceived by the academic community as representing a potential conflict of interest must be disclosed. If no such relationship exists, authors will be asked to confirm the following statement: 

The authors declare that the research was conducted in the absence of any commercial or financial relationships that could be construed as a potential conflict of interest.

\section*{Author Contributions}
%The Author Contributions section is mandatory for all articles, including articles by sole authors. If an appropriate statement is not provided on submission, a standard one will be inserted during the production process. The Author Contributions statement must describe the contributions of individual authors referred to by their initials and, in doing so, all authors agree to be accountable for the content of the work. Please see  \href{https://www.frontiersin.org/about/policies-and-publication-ethics#AuthorshipAuthorResponsibilities}{here} for full authorship criteria.
C.S. performed the MHD simulations and visualized the results. V.P. and K.K.R. performed IRIS, XRT, and SDO/AIA data analysis. B.C. contributed to spectrum prediction for IRIS. S.Y. and X.X. contributed to the synthetic SDO/AIA image modeling. All authors contributed to the model development, discussed the results, and commented on the manuscript.

\section*{Funding}
This work was supported by NSF grants AGS1723313 and AST2108438 to the Smithsonian Astrophysical Observatory.
C.S. acknowledges the support of NASA grants 80NSSC21K2044, 80NSSC19K0853, 80NSSC18K1129, and 80NSSC20K1318. V. P. acknowledges financial support from the NASA grants 80NSSC20K0716 and NNG09FA40C ({\it IRIS}). K.R. and X.X. are supported by NASA grant 80NSSC18K0732. S.Y. is supported by NSF grant AST-2108853 and NASA grant 80NSSC20K1283/SV0-09025 to NJIT.

\section*{Acknowledgments}
The computations in this paper were conducted on the Smithsonian High-Performance Cluster, Smithsonian Institution (https://doi.org/10.25572/SIHPC).
CHIANTI is a collaborative project involving George Mason University, the University of Michigan (USA), and the University of Cambridge (UK).
AIA data are courtesy of NASA/SDO and the AIA science team.
IRIS is a NASA small explorer mission developed and operated by LMSAL with mission operations executed at NASA Ames Research center and major contributions to downlink communications funded by ESA and the Norwegian Space Centre.

%\section*{Supplemental Data}
% \href{http://home.frontiersin.org/about/author-guidelines#SupplementaryMaterial}{Supplementary Material} should be uploaded separately on submission, if there are Supplementary Figures, please include the caption in the same file as the figure. LaTeX Supplementary Material templates can be found in the Frontiers LaTeX folder.

\section*{Data Availability Statement}
%The datasets [GENERATED/ANALYZED] for this study can be found in the [NAME OF REPOSITORY] [LINK].
% Please see the availability of data guidelines for more information, at https://www.frontiersin.org/about/author-guidelines#AvailabilityofData
The software used in this work are publicly available: Athena \citep{Stone2008ApJS..178..137S}, Astropy \citep{2013A&A...558A..33A}, SunPy \citep{sunpy2022zndo....591887M}.

\bibliographystyle{Frontiers-Harvard} %  Many Frontiers journals use the Harvard referencing system (Author-date), to find the style and resources for the journal you are submitting to: https://zendesk.frontiersin.org/hc/en-us/articles/360017860337-Frontiers-Reference-Styles-by-Journal. For Humanities and Social Sciences articles please include page numbers in the in-text citations 
\bibliography{mybib}

%%% Make sure to upload the bib file along with the tex file and PDF
%%% Please see the test.bib file for some examples of references

\section*{Figure captions}

%%% Please be aware that for original research articles we only permit a combined number of 15 figures and tables, one figure with multiple subfigures will count as only one figure.
%%% Use this if adding the figures directly in the mansucript, if so, please remember to also upload the files when submitting your article
%%% There is no need for adding the file termination, as long as you indicate where the file is saved. In the examples below the files (logo1.eps and logos.eps) are in the Frontiers LaTeX folder
%%% If using *.tif files convert them to .jpg or .png
%%%  NB logo1.eps is required in the path in order to correctly compile front page header %%%

%%% If you don't add the figures in the LaTeX files, please upload them when submitting the article.
%%% Frontiers will add the figures at the end of the provisional pdf automatically
%%% The use of LaTeX coding to draw Diagrams/Figures/Structures should be avoided. They should be external callouts including graphics.

\end{document}